# Thermodynamics of the Condensation of Dust Grains in Wolf-Rayet Stellar Winds

Anuj Gupta*, Sandeep Sahijpal
*Department of Physics, Panjab University, Chandigarh, India 160014*

**ABSTRACT**

Wolf—Rayet (WR) stars are the evolutionary phases of very massive stars prior to the final supernova explosion stage. These stars lose substantial mass during WN and WC stages. The mass losses are associated with diverse elemental and isotopic signatures that represent distinct stellar evolutionary processes. The WR strong winds can host environments for condensation of dust grains with diverse compositions. The condensation of dust in the outflows of the massive stars is supported by several observations. The present work is an attempt to develop a theoretical framework of thermodynamics associated with the condensation of dust grains in the winds of WN and WC phases. A novel numerical code has been developed for dust condensation. Apart from the equilibrium dust condensation calculations, we have attempted, perhaps for the first time, a set of non-equilibrium scenarios for dust condensation in various WR stages. These scenarios differ in terms of the magnitude of the non-equilibrium state, defined in terms of a simulation *non-equilibrium parameter*. Here, we attempt to understand the sensitivity of the simulation *non-equilibrium parameter* on the condensation sequence of dust grains. In general, we found that mostly C (graphite), TiC, SiC, AlN, CaS and Fe-metal are condensed in WR winds. The extent of non-equilibrium influences the relative proportions of earliest dust condensate compared to the condensates formed at later stages subsequent to the cooling of the gas. The results indicate that dust grains condensed in the WC phase may substantially contribute carbon-rich dust grains to the interstellar medium.

## 1 INTRODUCTION

The Wolf—Rayet (WR) phenomenon is generally associated with a single massive star with mass greater than 30 $M_\odot$, or, a comparatively less massive primary star, with mass greater than 15 $M_\odot$, in a close binary system. The phenomenon came into intense astronomical observation and discussion due to the unusual emission line spectra from the associated stars. These stars were initially considered as distinct due to their unusual emission lines unlike narrow absorption lines in the spectra of most of the stars. With the observational and theoretical advancements, it became clear that these rare stars are the evolutionary phases of very massive stars which have almost reached the end of their nucleosynthesis processes prior to the final supernova explosion stage (see e.g. Woosley, Heger & Weaver 2002; Sahijpal & Soni 2006; Meynet et al. 2017). The rareness of these stars is because of the less production of these massive stars and their short evolutionary life-span in the WR phase (~ a few $10^5$ years). According to the Salpeter's Initial Mass Function (IMF), there are only 3 stars that are born with the mass more than 8 $M_\odot$ corresponding to the birth of 1000 stars in a single stellar cluster. Even though these massive stars form only about 14 per cent of the total mass of a single stellar generation, their contribution to the chemical enrichment of the interstellar medium (ISM) is significant especially during their substantial mass loss through stellar winds (Meynet et al. 2017).

Based on the spectroscopic analysis of the emission lines, WR stars can be broadly divided into three classes, viz., WN, WC, and WO (Torres, Conti & Massey 1986). During the WN stage, the emission lines of helium and nitrogen dominate the spectra as the massive stars eject their H-rich envelope and are left with He- and N- excess. Although the products of the CNO cycle are deficient in H, C and O, hydrogen and carbon emission can be seen in a few WN stars. In contrast to the WN phase, H-emission is completely absent from the spectra of WC stars which displays the products of He-burning. During the WC/WO stage, the He-rich envelope is removed by strong stellar winds and the stellar spectra indicate the presence of the C- and O- excess. In the case of a binary system, the envelopes are stripped off through Roche lobe overflow (Prantzos 1984; Icko & Livio 1993). Finally, the WR stars are left with the bare CO core that eventually collapses as SN Ib/c. Both the WN and WC phases are further divided into early and late-type subclasses. In case of WN stars, early WN phase (WNE) exhibits emission lines of highly ionized species (WN2—5), whereas, the stages, WN7 to WN9(—11), associated with low-ionized species, are known as late WN phase (WNL). Notice that the early or high ionization stage appears after the late-type evolutionary stage. In general, H emission is shown only by WNL stars and not by the WNE stars, though the exceptions do exist. Similarly, the high-ionization stages WC4—6 and the low-ionization stages WC7—9 are known as early (WCE) and late type (WCL) WC stars, respectively (Crowther 2008). The heated dust is in general found in the surrounding of low-ionization stages of WN and WC stars.

* E-mail: mr.anuj@pu.ac.in

The mass losses induced by stellar winds and stellar rotation are the most important features which significantly influence the evolution of WR stars (Langer 2012). The mass-loss rates during the initial stages of evolution in turn strongly affect the subsequent evolutionary phase of these massive stars (Smith 2014). The rotation also affects the evolutionary track through mixing processes and induces anisotropies in the winds (Heger & Langer 2000). These effects collectively decide the chemical composition of the outflow that in turn determines the chemistry of the ejected shell (Meynet, Maeder & Hirschi 2003; Maeder 2009). For instance, a 60 $M_\odot$ star evolves through the late-type WN stage when the stellar mass becomes approximately 35 $M_\odot$. He, H, N are the three most abundant elements during this phase. During the WNE phase at approximately 30 $M_\odot$ mass, hydrogen is totally lost from the envelope. He, N and C become the three most abundant elements in the updated composition of the stellar winds during this phase. He, C and O become the most abundant elements during the next evolutionary stage, WC, that appears approximately at 25 $M_\odot$ mass. The distinct compositions favour a wide range of dust grains condensation. We present the condensation sequence of dust grains in the outflows of the massive stars.

Dust grains are the key ingredients of the baryonic matter distribution in the ISM as the bulk abundance of the non-volatile and moderately volatile elements are trapped in these grains. The volatile elements mostly reside in the form of interstellar gas. The dust plays a major role in terms of the formation of molecules in the interstellar medium (Gould & Salpeter 1963) that in turn influences star formation within the interstellar molecular clouds. The astronomical observations and studies of the interstellar dust grains separated from meteorites in laboratory have indicated the presence of dust grains with a wide range of chemical and isotopic composition that in a manner serve as a marker of various stellar nucleosynthetic processes occurring in various astrophysical environments (Amari et al. 1993; Nguyen, Keller & Messenger 2016). These environments differ in terms of pressure, temperature and chemical composition, and hence, provide distinct dust condensation scenarios. To understand the nature and the composition of dust grains, it becomes extremely important to deduce the thermodynamical criteria associated with the condensation of grains with distinct compositions.

Dust condensation requires dense regions at low temperatures like the circumstellar envelopes of low- to intermediate-mass stars in the late stages of their evolution, and in the ejecta of supernovae where suitable conditions for condensation can be achieved (Zinner 1997). The interstellar dust growth generally commence with the condensation of carbon or silicate rich grains in the circumstellar environments which later on accumulate additional moderately volatile and volatile elements to form icy mantles of water ice, methane, carbon monoxide, and ammonia, depending upon the astrophysical environment (Saslaw & Gaustad 1969; Fraser, Collings & McCoustra 2002). The various factors affecting the condensation sequence are the prevailing pressure, temperature and elemental abundance that determines the partial pressure of various elements in a gas (Ebel 2006). Several observational pieces of evidence support the dust condensation in the outflows of WR stars (Williams et al. 1985, 1992, 2001; Williams 2014). The present work is an attempt to develop a comprehensive theoretical framework of thermodynamics associated with the condensation of dust grains in the winds of Wolf—Rayet stars using equilibrium as well as non-equilibrium condensation scenarios.

In order to investigate the possibility of the condensation of dust grains in the winds of Wolf—Rayet stars, several WR stars such as WR125, WR137, and WR140 were earlier studied by other workers, and the dust was found to be condensed, with an estimated mass of about 0.5 per cent of the mass-loss rate (Williams et al. 1985, 1992). It was indicated that there is an episodically dust making class of WR stars (Williams et al. 2001). At first glance, the dust condensation in such harsh environments seems to be surprising, however, there exist conditions under which dust condensation is feasible (Usov 1991). During the compression of gas in the stellar winds, heat is generated and radiated out, thereby, cooling the gas for the condensation of dust grains. This episodic dust formation has been observed in the WR stars in binary systems. Binarity enhances the wind density required for dust formation. If WC 7-9 type star is in a binary system with another O type star, the collision of their stellar winds at the periastron provides the conditions in which condensation is likely to happen. As the system moves through periastron passage in an eccentric binary orbit, the stellar winds interact, and their compression releases the energy through radiation that eventually cools the region sufficiently, thereby, resulting in the condensation of dust grains (Crowther 2003). WR stars are not the only astrophysical environments where episodic dust formation occurs. There are various types of astrophysical environments allowing such episodic events. Novae are the closest analogue to WC stars in context to dust formation. Dust condensation in classical novae has been discussed in detail in earlier studies (Bode 1995) which emphasized the clumping of stellar ejecta as an essential requirement for the nucleation and growth of grains. Clumping of stellar winds enhances the density in localized pockets which eventually allows the dust to condense in the circumstellar environment. In the case of WR stars, the density enhancement by a factor of 4 is in good agreement with WN spectral features, whereas, in the case of WC class stars, the density enhancement by a factor of 16 produces a better observational match (Hamann & Koesterke 1998).

There are observational pieces of evidence of persistent dust maker WR stars (Williams 1995). The conditions that produce high-density regions for the persistent dust makers are still not clear. However, clumping in the stellar winds, disk structure associated with the winds, and compression Colliding Wind Binary (CWB) are the likely conditions for dust condensation. The WC 8-9 stars, e.g., WR 104 and WR 106, are known to be prominent dust makers, despite not being the members of CWBs. Isolated clumps in stellar winds are another class providing high-density structure required for dust formation (Williams 2014). However, the exact phenomenon taking place in these stars is still an open problem. The present work is a theoretical attempt to explain the observations and understand the nature of the environment and processes occurring therein.

Thermodynamical equilibrium calculations provide an insight into the chemical nature of the dust that can condense in a gaseous system cooling over time. The equilibrium condensation calculations assume that dust gets sufficient time to condense as the gas cools down slowly, and actively interact with the remaining gas in terms of the further chemical reaction. In the inner regions of the stellar envelope, the temperatures and pressures are very high, and the time taken by a typical thermochemical reaction is relatively less than the cooling time. There is a possibility that some of the SiC grains were produced in such a region where thermochemical equilibrium prevails (Amari et al. 1995). There are observational pieces of evidence for dust formation in η Carinae which is thought to be one of the most luminous and massive stars in the Milky Way. The dust, which was probably formed during the historic 1843 outburst, has been observed in the ejected shells of η Carinae. Thermodynamic equilibrium calculations of η Carinae predicted the possibility of metallic iron, forsterite and SiC grains (Gail, Duschl & Weis 1999). The condensation of amorphous carbon through the clustering of carbon atoms has been proposed with the details of the chemical pathways for the production of soot particles (Cherchneff & Tiellens 1995). Zubko (1995) concluded that the possibility of condensation of graphite grain is more than the amorphous carbon. Amorphous carbon can stick as a mantle on to the core of graphite grain. There is a possible link of this condensation with the broad anomalous 2200 Å absorption feature that is caused by graphite grains. Earlier studies attempted to find the link but were not able to establish it (Williams et al. 1985). However, it cannot be ruled out. Further, the isotopic measurements of meteorites show isotopic heterogeneity of the solar nebula that is believed to be caused by presolar grains. Presolar grains are believed to have been produced in stellar outflow, novae and supernovae ejecta (Nittler 2003). It is widely believed that AGB stars and SNe are the potential sources of the graphite and carbide presolar grains. However, Wolf—Rayet stars may be possible candidates for introducing their condensed dust particles in the presolar cloud (Amari, Zinner & Lewis 1996; Arnould, Meynet & Paulus 1997; Nittler et al. 1997; Gaidos et al. 2009). It is well established that meteorites contain presolar C-bearing grains (Anders & Zinner 1993). In the present work, we explore the possibility of dust condensation in the WC phase that may have contributed to these carbon and carbide presolar grains present in the early solar system.

Although most of the theoretical studies dealing with dust condensation assume thermodynamical equilibrium, the astrophysical environments can, in general, be in a non-equilibrium state. For instance, the thermodynamic equilibrium does not prevail in the winds of WR stars as proved by earlier works (Donn & Nuth 1985; Cherchneff 2010). The basic parameters governing thermodynamic equilibrium are, i) the temperature that determines thermal equilibrium, ii) the pressure that defines the mechanical equilibrium, and iii) the chemical potential that determines the chemical equilibrium. In the case of a non-equilibrium state of matter, the pressure and number density continuously change as the system evolves in time due to expanding WR wind fronts. The effective number density may also change due to the removal or isolation of condensed dust from the system. The isolation may happen in terms of grain growth or coagulation of dust particles with cooling, thereby, resulting in the formation of a larger dust grain at the cost of several smaller dust grains. This effectively reduces the surface area to volume ratio of the dust grains that can hamper an efficient interaction of condensed grains with the residual gas, thereby, resulting in a non-equilibrium condition. The non-equilibrium situation can also be caused by differential velocity between gas and the condensed dust on account of gas drag or radiation pressure within the WR winds. The radiation pressure provides a drift velocity to the grain relative to the wind in which it condensed. The drift velocity depends upon the density of the winds. In isotropic WC winds, this velocity can be ~ 100 km $s^{-1}$ whereas it will be much smaller for compressed winds (Williams et al. 2001; Zubko 1998). Thus, the grain growth and differential velocity between gas and dust collectively decide the fraction of dust, which got isolated from the gaseous system in terms of its active chemical interaction with the residual gas. The removal of dust affects the equilibrium of the system. For instance, the sticking of amorphous carbon on the core of graphite grain isolates the core part from the surroundings. This reduces the effective number density of reactive species in the system and thus leads to chemical non-equilibrium. We have made an attempt in the present work, perhaps for the first time, to numerically simulate the thermodynamical non-equilibrium state for dust condensation in the WR environment. This is achieved by invoking a numerical non-equilibrium thermodynamical parameter that can be varied to understand the extent of the non-equilibrium condition, and its influence on the sequence of grain condensation.

The theoretical technique involved in thermodynamical modelling is discussed in section 2. The method to compute the chemical equilibria is explained in subsection 2.1. Non-equilibria formalism is explained in subsection 2.2, with the technical details described in subsection 2.3. The details of the elemental abundance data, clumping factor, pressure values and the thermodynamic data of all the considered species are discussed in subsection 2.4. The results of our simulations are presented in section 3. Section 4 includes the discussion of the wide-range of condensation scenarios considered in the present work. Finally, the main conclusions drawn from the present work are mentioned in section 5.

## 2 METHODOLOGY

### 2.1 Method of calculation

In the present work, we have adopted the thermodynamical approach to study the condensation of dust grains. The thermodynamics provides comprehensive results for the condensation calculations and distribution of matter in a multi-phase chemical system. The conventional approach to model the thermodynamic equilibrium condensation calculations is to solve the non-linear mass balance equations involving all the considered gaseous elements simultaneously with the mass action equations of the condensed species (Grossman 1972; Lattimer, Schramm & Grossman 1978; Ebel et al. 2000). We assume that the basic mechanism of solving the equations remains the same even in the case of non-equilibrium thermodynamical scenarios. A set of twenty abundant elements (H, He, C, N, O, F, Na, Mg, Al, Si, P, S, Cl, K, Ca, Ti, Cr, Fe, Co, and Ni) has been considered in the present study. The program involves 20 mass balance equations corresponding to each element, similar to equation (1).

$$N_i^{\mathrm{tot}} = \sum_i v_{ij} * N_j \qquad (1)$$

Here, $N_i^{\mathrm{tot}}$, is the total number of moles per liter of the $i^{th}$ element, $N_j$, is the number of moles per liter of the $j^{th}$ gas species that can be formed by its constituent monatomic elements, and $v_{ij}$ is the corresponding stoichiometric coefficient of the $i^{th}$ element in $j^{th}$ gas species. $N_j$ can be written in terms of the partial pressure of the species '$j$' that is determined using partial pressures of its basic elements '$i$' as given in equation (2).

$$P_j = K_j \cdot \prod_i^{elements} P_i^{v_{ij}} \qquad (2)$$

Here, $K_j$ is the equilibrium constant of reaction for the formation of product gas species from its basic elements. These equations are solved simultaneously to achieve the point of convergence of the system at an initial temperature. The gas phases abundances are solved iteratively with the decreasing temperature, and the values of the partial pressures of the elements are calculated at each falling temperature step. In order to reduce the calculation time at each temperature step, the process is continued by using the results of current calculation as the initial state for the next iteration. The calculated partial pressures of the elements are used to assess the stability of the solid species. The stability of any condensate in the pure phase is assessed by comparing its Gibbs energy value with the Gibbs energy of formation calculated using the partial pressures of the elements of monatomic gaseous species from which that condensate is formed. The stability criterion is given by equation (3).

$$\left(\prod_i^{elements} P_i^{v_{ij}}\right)_{pr} \geq A_j K_j \qquad (3)$$

Here, the left-hand side of the equation gives the value of equilibrium constant using the partial pressures calculated from the program, and $A_j$ represents the activity of the condensate. The activity is one for the pure phase and is estimated by the product of the mole fraction and activity coefficient for a solid solution. We have considered only the ideal solid solutions. The activity coefficient is taken as unity in an ideal solid solution. The condensate is added to the system if it is found to be stable. This leads to the addition of a mass action equation (4) of that condensate in the system.

$$logK_j = v_{ij} * \sum_i^{elements} \log P_i - logA_j \qquad (4)$$

The set of mass balance equations is solved simultaneously with the mass action equation. The occurrence of grain nucleation is still an open problem. The gas to dust phase transition has been taken directly as a function of temperature based on the minimization of Gibbs free energy of the assemblage. We iterate the calculations to successively lower temperature for a bulk system composition at various total pressures.

When a new condensate appears in the solid phase, the corresponding mass of its constituent elements is subtracted from the mass balance equation. The numerical process of appearance of distinct condensates continues as the thermodynamics allows the system to evolve with decreasing temperature. In a pure equilibrium system, once a condensate appears in the system it reacts with the gas as well as with the other condensates in the assemblage, and thus, forms new species which can acquire stability at some temperature according to their thermodynamical value. The appeared condensate is removed from the system when it is found in a vanishingly small amount. An equivalent amount of the corresponding mass is added back to the gas reservoir to maintain the mass balance. The whole process is repeated for the modified assemblage in which the matter is redistributed among the available phases.

To characterize the cooling rate of the gas associated with WR winds, we assume an exponentially reducing temperature profile as described in the equation (5).

$$T\,(t) = T_0\, e^{-t/\tau}, \qquad \text{for } t \geq 0 \tag{5}$$

Here, $\tau$ is the characteristic time-scale which can be a few days or a few 100 days. The characteristic time-scale is chosen such that clumped gas pocket associated with WR wind moved sufficiently away from stellar radiation, and the temperature dropped substantially to cause grain condensation. Further, we assume that the dust survives in a solid phase due to geometrical dilution of the stellar radiation heating, far away from the WR star. $T_0$ is the initially assumed temperature at which the first solid condense in a given environment. If a system is cooling down slowly, it will have a large characteristic time. This can be realized as an equilibrium case because the relaxation time is less than the cooling time. On the contrary, if characteristic time is less, the species do not get enough time to interact in the system. This results in a non-equilibrium scenario as the inactive or isolated dust species no longer participates in dust-gas equilibrium chemical interaction.

The temperature drop, in a manner, represents the time evolution of the cooling gas according to the equation (5). Hence, the time evolution in the present work is considered in terms of a drop in the temperature from an initial temperature, $T_c$. corresponding to the initiation of the dust condensation.

## 2.2 Non-equilibrium Calculations

We have also incorporated non-equilibrium thermodynamical states of matter along with the equilibrium states. We have considered distinct scenarios to deal with non-equilibrium chemistry in various phases of WR environments. In the first scenario, a complete equilibrium between dust and gas has been thoroughly studied. The system cools down slowly at an assumed constant pressure. In this scenario, the system remains in complete thermodynamic equilibrium at every step. The second scenario involves the gradual change in pressure with the cooling of gas because of the expulsion of WR winds from the star. We simulated the system by considering the pressure as a temperature-dependent variable in an adiabatic manner (using $\gamma = 5/3$ in the equation, $PT^{\frac{\gamma}{1-\gamma}} = Constant$). This is a kind of mechanical non-equilibrium scenario.

We have also numerically simulated three distinct chemical non-equilibrium scenarios by introducing a *numerical non-equilibrium parameter*, *f*, that deals with the extent of the fraction of already condensed dust that will be subsequently removed from equilibrium with the residual gas for any further thermodynamical condensation considerations. This fraction of the condensed dust will be gradually removed from gas-dust equilibrium with the step-wise lowering of temperature, thereby, enabling us to understand the nature of non-equilibrium condensation. As mentioned in the introduction, the probable reasons for the assumed gas-dust non-equilibrium state could be either due to the grain growth that hampers the inner regions of the grain to be in equilibrium with the gas or anticipated partial segregation of dust and gas due to differential velocity experienced in clumped WR winds. In the present work, we could explore the non-equilibrium effects only in terms of varying pressure and the dust-gas partial equilibrium conditions. The numerical approach of our adopted technique to invoke a non-equilibrium condition is presented in the following. However, we present a semi-analytical approach for non-equilibrium calculations based on our assumptions before the adopted numerical approach. In the semi-analytical approach, we assume that $N_e$ is the amount of an element, say X, present in the gas phase at any temperature T. With the decrease in temperature, the element condenses into a solid species. The change in the abundance in the gas phase with the change in temperature depends upon the abundance of an element at that temperature. Mathematically,

$$\frac{dN_e}{dT} = K_e N_e \tag{6}$$

$K_e$ is a proportionality factor, and it determines the rate of change of $N_e$. It is assumed to be a constant, even though, it could be a function of temperature. This assumption will not influence our adopted numerical approach, whereby, we piecewise handle the non-equilibrium thermodynamics at every temperature drop in a systematic manner. A positive sign is taken because $N_e$ is decreasing with a decrease in temperature. Integrating, the equation (6) within the limits, we get an approximate solution.

$$\int_{N_{e0}}^{N_e} \frac{dN_e}{N_e} = \int_{T_c}^{T} K_e dT \tag{7}$$

Here, $N_{e0}$, is the initial abundance of the element in the gaseous state at the earliest condensation temperature $T_c$ where the element, X, begins to condense. The equation (7) yields an analytical solution, presented in equation (8), for the abundance of the element, X, in the gas phase at any falling temperature step.

$$N_e = N_{e0}\, e^{K_e(T-T_c)} \tag{8}$$

The element, X, can be a constituent of several condensates. The mass of condensate increases with the decrease in temperature as the condensation process proceeds. A condensate can further react with the gas, and thus, its abundance can decrease with the formation of new condensates. The change in the amount $(N_c)$ of the element, X, in condensate on account of condensation as well as conversion to another distinct dust species can be written as,

$$\frac{dN_c}{dT} = -f_c K_e N_e + K_c N_c \tag{9}$$

The first term here represents the condensation of an element in one specific dust species, whereas, the second term deals with the conversion of the dust species into another form on account of the gas-dust chemical reaction. $K_c$ is a constant proportionality factor. It determines the rate of change of the first condensed species on account of conversion to another dust species. $f_c$ is the fraction of the element, X, that condenses into the first condensate, and its value is $\leq 1$ as one element can condense in more than one condensate. However, if we assume the conversion into only one another condensate, we can obtain a simple analytical solution for the equation (9). We assume that $N_c = 0$ at $T = T_c$. Integrating and solving equation (9), we get

$$N_c = \frac{-f_c K_e N_{e0}}{(K_e - K_c)}\left[e^{K_e(T-T_c)} - e^{K_c(T-T_c)}\right] \quad (10)$$

The intermediate steps have been given in supplementary file. The condensate abundance decreases with the formation of other dust species. Further, we assume that some fraction of the condensate always gets isolated from the gaseous system in terms of gas-dust non-equilibrium, as mentioned earlier. This non-equilibrium content of the condensate gradually increases according to the abundance of the condensate. The gradual increase in the non-equilibrium fraction of the element, X, due to the segregation of the dust from the gas component can be described in terms of $N_{ic}$, according to equation (11).

$$\frac{dN_{ic}}{dT} = -f_{ic} K_c N_c \quad (11)$$

Here, $f_{ic}$ is the part of the condensate that got isolated from the gaseous system due to the non-equilibrium dust-gas system. The equation (11) can be analytically solved using the initial abundance, $N_{ic}$, of the condensate for the element, X, obtained from the equation (10).

In principle, the abundances of all the elements, as well as their corresponding condensates, follow the above formalism. Here, $f_c, f_{ic}, K_e, K_c$ are assumed to be the proportionality constants that are different for different elements and their respective condensates. These parameters could be functions of temperature as the processes associated with grain condensation and chemical reactions involving gas and dust are temperature dependent. This delimits the use of the semi-analytical approach in developing the non-equilibrium condensation formulation. In the following, we present the numerical approach that does not suffer from this limitation of the semi-analytical approach.

## 2.3 Computational details

A novel code in Python has been developed that investigates the mineralogical condensation sequence in various phases of WR stars. The mass balance equations of all the twenty considered elements and mass-action equations of the condensed phases are solved iteratively to yield the simultaneous solutions for the system. The equations involved are highly non-linear in nature. These are efficiently solved using a modified Powell method (Chen & Stadtherr 1981). A strict chi-square test ensured global convergence. The order of the tolerance for convergence was set to 20. Solving such equations is not scale-invariant, so it should appropriately be normalized to increase precision. The temperature step has been kept dynamic in the program. It has been taken to be 1 K in general, but whenever the system raises an error for this step-size or a new condensate becomes stable or some condensate disappears, the temperature step is reduced automatically. The program then evolves the system at a temperature step of 0.1 K. The smaller temperature step resolves the condensation sequence of multiple phases as multiple events cannot take place in this small step size.

As mentioned earlier, in order to incorporate the non-equilibrium scenario, the semi-analytical approach involves several unknown factors like $f_c, f_{ic}, K_e, K_c$. Since the values of so many unknown parameters are not available, it is not feasible to obtain the absolute abundances using this approach. The possible temperature dependence of these parameters further complicates the issue. Therefore, we adopted a numerical approach to handle the problem. The abundances of condensates are calculated through the program using the thermodynamical approach. A constant fraction of the condensate is isolated from gas-dust equilibrium from the system at every temperature step to deal with the problem in a chemical non-equilibrium sense. Out of the total inventory of the dust which condenses in the system assemblage, a fraction $f_{ic}$ is separated from the gas at every step during the decline in temperature. This fraction does not take part in any chemical reaction afterward. The remaining fraction of the dust, $f = (1 - f_{ic})$, which remains in equilibrium with the gas, reacts to form other condensates in the system. As already explained in the introduction, $f_{ic}$ could be treated as a function of grain growth rate and the differential velocity between gas and dust within the WR winds. The removal of a dust fraction decreases the effective number density of the condensate as well as the constituent elements. This modifies the system assemblage in terms of abundance ratios that affect gas grain chemistry. In the numerical code, the equilibrium fraction, $f = (1 - f_{ic})$, of the dust, is kept in mass balance constraint, and the associated elemental number density is modified according to the isolated fraction of the dust at every temperature step corresponding to 1 K fall. The fraction, $f_{ic}$, has been varied from 0 to 0.05 of the condensed dust at every step during the decrease in the temperature. The fraction, f, is treated as a simulation *non-equilibrium parameter*.

As mentioned earlier, the fraction, $f_{ic}$, could be accounted for by the gradual reduction in the surface area to volume ratio due to grain growth. We have considered an analysis of a spherical grain growth initiating from a 1 $\dot{A}$ size to a final radius of 1 µm. The surface area to volume ratio of the grain declines from 0.3 to $3\times10^{-4}$, thereby, hampering the participation of the inner regions of the grain to be in equilibrium with gas. We assume that the grain linearly grows in terms of concentric annular spheres such that the radius increases in a linear manner with the gradual fall in the ambient temperature. If we assume that the 1 µm is formed over a temperature drop of 500 K, divided equally into temperature steps of 1 K, the fraction, $f_{ic}$, of the condensed grain that gets buried inside the grain against equilibrium with gas varies steadily from 0—0.05 until the grain acquires 90 per cent of its final size. Thereafter, the fraction increases sharply till complete growth of 1 µm sized dust grain. In this analysis, we have assumed that the entire matter for the grain growth is available right from the onset time of the grain growth. We have performed four sets of simulations with an assumed constant fraction, $f_{ic}$, of 0, 0.005, 0.01 and 0.05, throughout the simulations. This will correspond to the fractional values ($f$) of 1, 0.995, 0.99 and 0.95, respectively, of the condensed dust that remains in equilibrium with the gas for further chemical reactions at every drop in the temperature by 1 K. If we consider an initial '*x*' contents of a condensate at a specific initial temperature, $T_c$. For 100 K temperature drop, with a 1 K temperature step, this will correspond to the final dust contents of, *x*, ~0.6 *x*, ~0.37 *x* and ~0.006 *x*, that are in equilibrium with the gas for the fractional values ($f$) of 1, 0.995, 0.99 and 0.95, respectively. Our simulations indicate that any further decrease in the fraction, $f$, does not substantially influence the major conclusions drawn from the present work. It should be mentioned that the differential velocity introduced among dust and gas species within the WR winds can cause further enhancement in non-equilibrium conditions within dust and gas. As discussed in the following, the non-equilibrium effects can be seen more predominantly in the case of simulations with 0.05 fractional removal of dust from equilibrium at each reducing temperature step. The behavior of the system can be anticipated for any value of the dust-gas equilibrium ratio. It can be extrapolated or interpolated from the effects visible for the considered values. It should be noticed that we have incorporated dust-gas non-equilibrium only. The gas-gas non-equilibrium has not been incorporated, as that can dramatically alter the elemental composition of the system assemblage. Hence, the system will not remain identical to the considered WR environment. We do not assume such a situation in the case of the WR environment.

Irrespective of the nature of a simulation, the program compiles all the intermediate output results after completion and yields the final numbers and the mole fractions of all gaseous and solid condensed species in Excel format that are ready for plotting.

## 2.4 Parameters, calculation and thermodynamic data

The composition of the considered environment is required in thermodynamical calculations. The nucleosynthetic yields were estimated for the non-rotating (initial rotational velocity = 0 km $s^{-1}$) as well as rotating (initial rotational velocity = 300 km $s^{-1}$) stellar models for the WN (WN_NR and WN_R) and WC (WC_NR and WC_R) phases of a 60 $M_\odot$ WR star from the available mean enhancement factors of various nuclei (Meynet et al. 2001). The yields so obtained also incorporate important effects like mass-loss rates, clumping of the stellar winds, etc. We have considered a 60 $M_\odot$ mass WR star to represent an average of WR population. Along with this approach, the abundance has also been obtained from the mass fraction of the various nuclei during different stages of the evolution of 60 $M_\odot$ massive stars of solar metallicity (Maeder 2009). The abundances have also been estimated for the late and early phases of WN stars. The three most abundant elements in the WN and WC phase are H, He, N and He, C, O, respectively. H is taken to be zero in WNE as well as the WC phase. It should be noticed that radiation gas has not been considered in the simulations. Although the ionizing radiations or the ions in the system assemblage can produce significant changes in the physicochemical processes occurring in that environment, the numerical code developed in the present study has the limitation to deal with the plasma physics.

The identity of circumstellar dust merges with the interstellar dust at a distance of 1 parsec from the central star (Evans 1993). If we take an average velocity of 1000 km $s^{-1}$ of the stellar winds, it will take approximately 977 years to reach such a large distance from the star. The observations of WR stars support much earlier condensation of dust. It is difficult to determine the plausible range of density and pressure that can reliably be used in thermodynamical calculations. We have assumed a spherically symmetrically expanding WR envelope in which small scale structuring of winds occurs at a distance of a few thousand of stellar radii due to inhomogeneities or instabilities in the stellar winds. The stellar winds coming from the WR phase move outwards into the surrounding environment at a few ~1000 km $s^{-1}$ terminal velocity. The velocity decreases in the outer zones and a velocity gradient come into action. The clumping of winds takes place at certain places. The density value has been chosen high (~ $10^{13}$ $cm^{-3}$) at an approximate distance of ~ $10^{15}$ cm from the star to explore the possibility of dust condensation (Cherchneff et al. 2000; Usov 1991). The density value is even higher at the periastron due to the colliding winds of binary stars. The clumping causes density enhancement. In the present work, the density enhancement has been taken to be 10 and 25 to cover the complete range. The corresponding density values give the pressure value ranging from $10^{-6}$ to $10^{-3}$ bar. Various phases of WR stars have been simulated in this pressure range for grain condensation.

The thermodynamic data library is an essential input component of the above-mentioned technique. We assembled a library of thermodynamic values of all the considered species. The details of all the gaseous and solid species considered in the present study and their thermodynamical sources are presented in Table 1 (Supplementary file). The thermodynamic data have been taken from the standard compilations (Barin 1995; Chase 1998). The JANAF database was downloaded by the program in the text format from the National Institute of Standards and Technology. There was a typographical mistake of a negative sign in many text files, which was cross verified from the data in the pdf version available on the same source. The typo was corrected in all the files manually and the final corrected data values were fetched by the program automatically. There is an error in the JANAF database for gaseous HS (Pasek et al. 2005), and hence, it was taken from Barin (1995). Also, the data for the P-bearing compounds PH, $PH_3$, and PN in the gaseous phase are not correct due to the revision in the 'P' reference state in JANAF fourth edition. This data was taken from Lodders (1999). Except for these species, the data for all other gas phase were taken from JANAF thermochemical tables (1998). The data for all the liquid species were also taken from JANAF thermochemical tables. The data of pyrope, silicon oxynitride, and Ca-aluminates have been taken from other sources ( Fegley 1981; Kumar & Kay 1985) as mentioned in Table 1.

**Table 1.** Library of all the considered species and sources of their thermodynamic data.

| **Gas Species** |
|---|
| Al, $Al_2$, $Al_2Cl_6$, $Al_2F_6$, $Al_2O$, $Al_2O_2$, AlC, AlCl, $AlCl_2$, $AlCl_2F$, $AlCl_3$, AlClF, $AlClF_2$, AlF, $AlF_2$, $AlF_3$, AlH, AlN, AlO, $AlO_2$, $AlO_2H$, AlOH, AlS, C, $C_2$, $C_2Cl_2$, $C_2Cl_4$, $C_2Cl_6$, $C_2F_2$, $C_2F_3N$, $C_2F_4$, $C_2F_6$, $C_2H$, $C_2H_2$, $C_2H_4$, $C_2H_4O$, $C_2HCl$, $C_2HF$, $C_2N$, $C_2N_2$, $C_2O$, $C_3$, $C_3O_2$, $C_4$, $C_4N_2$, $C_5$, Ca, $Ca_2$, CaCl, $CaCl_2$, CaF, $CaF_2$, CaO, $CaO_2H_2$, CaOH, CaS, …. |

* Here, only the first few rows are given. The full Table is available online in supplementary data as Table S1.

## 3 RESULTS

We have run dozens of simulations for estimating the temperature of appearance (and disappearance) of condensates with a varied composition of distinct Wolf-Rayet stages. The simulations were performed with equilibrium as well as a range of non-equilibrium thermodynamical scenarios by varying the pressure and the fractionation of condensed dust to gas. Two distinct sets of simulations were performed with initial total pressures of $10^{-3}$ and $10^{-6}$ bar. The total pressure of the system assemblage has been assigned a static as well as a dynamical value for the simulations. In the static case, constant pressure has been maintained throughout the simulation, whereas, in the dynamic case, the adiabatic expansion has been assumed by varying the pressure. We have considered six different WR compositions, with distinct dust to gas equilibrium fractions, $f$, ranging from 0.95 to 1 for the system evolving in an adiabatic manner. The fraction, $f$, is treated as a simulation *non-equilibrium parameter*. The abundances for the different stages of WR stars have been normalized with respect to silicon. The relative elemental abundances present in the assemblage and the total pressure of the system collectively decides the partial pressures of the elements. The mineralogical condensation sequence has been studied for all the models.

The variations in the condensation temperature have been studied for WN and WC phases for non-rotating as well as rotating stellar models. WN late and early phases have also been explored. The *normalized* mass distribution of the condensed phases and gas phases for distinct WR compositions are shown in Figs. 1-10. Figs. 1, 2, 5-10, representing the normalized mass distribution of condensates, contain five panels with distinct adopted non-equilibrium thermodynamical criteria. In the non-equilibrium condensation scenarios, the normalized mass included both the equilibrium as well as non-equilibrium dust components that have been removed from interaction with the residual gas. The panel (a) in these figures represent full dust to gas equilibrium thermodynamical scenario along with constant total pressure. The subsequent panels (b-e) represent the scenarios involving dynamic pressure by considering the adiabatic variation of pressure with the reduction in the temperature of the expanding WR winds. The panel (b) represents dust to gas equilibrium scenario with dynamic pressure. The results in this scenario are almost identical to the scenario (a) but the condensation temperatures are reduced. The panel (c) of the figures represents the scenario in which 0.995 fraction, $f$, of the dust remains in equilibrium with the residual gas at any temperature step of 1 K during the cooling. The panel (d) represents the dust to gas equilibrium fraction $f$ of 0.99. The panel (e) represents the maximum deviations from the equilibrium considered in this work, and only 0.95 fraction of the dust remains in equilibrium with the gas at a given temperature step of 1 K during its cooling. The normalized mass distribution of gas-phase has been presented in the case of WN and WC compositions for the non-rotating stellar model corresponding to 0.95 dust to gas equilibrium fraction at $10^{-3}$ bar total pressure (Fig. 3 & 4). The normalized mass distribution for non-rotating WC composition and late-type WN composition at $10^{-6}$ bar pressure has been presented in Figs. 9 & 10. The results obtained from various simulations have been tabulated in the form of temperatures of appearance (and disappearance) of various condensates for all the compositions (see Tables 2a-9a). The condensate names have been mentioned for solid-solutions, whereas, the pure solid phases have been written by their formulas in tabular representation. The various columns in Tables 2a-9a represent the name of the condensates which become stable with decreasing temperature and the condensation temperature. The condensation temperature represents the value of temperature at which the condensed phase stabilizes for the first time in a given composition. Along with this, peak value of normalized mass distribution of stable condensates in simulated temperature range have also been given (see Tables 2b-9b). The various columns in Tables 2b-9b represent the name of the condensates which become stable with decreasing temperature and the maximum value of normalized mass which the condensates attained at certain temperature. The results of WN, WC phases in non-rotating star have been presented in Tables 2 and 3 for all the considered dust to gas equilibrium conditions. Tables 4 and 5 represent the results of WN and WC phases for a rotating stellar model. The condensation sequence for WNL and WNE compositions at $10^{-3}$ bar pressure are given in Tables 6 and 7. The models corresponding to the WNL phase and WC compositions for rotating star at $10^{-6}$ bar pressure are tabulated in Tables 8 and 9. Column 1 represents the names of the condensates and the column 2 to 6 represent the corresponding condensation temperatures in Tables (a) and peak mass value in corresponding Tables (b) at dust to gas equilibrium fraction used in the present work.

We assumed a completely homogenized and vaporized environment at the initial stage of the simulation with distinct phases of WR stars at various pressures. Before the onset of stability of any solid phase, aluminium mostly remained in monatomic Al form and a minuscule part of it remained in AlC, $Al_2O$ and AlH forms. Calcium, magnesium, titanium, and iron mostly remained in their monatomic form. Carbon and oxygen primarily formed CO molecules. CS, $C_3$, HCN, $SiC_2$, SiO, SiS and TiO are the other gaseous species that became prominent at different stages of the simulation according to the composition. For instance, in the static case of WC star, ~36 per cent of Si is in $SiC_2$ molecule at $10^{-3}$ bar pressure. This percentage decreases and approaches 0 at a pressure of $10^{-6}$ bar. On the other hand, WN stellar environment provides the stability to SiS, SiO, and TiO like species, and a significant amount of their mole fraction exists in the gaseous phase. The results have been discussed in detail in the discussion section.

## 4 DISCUSSION

An attempt has been made in the present work to investigate the condensation of dust grains in the environment of Wolf-Rayet stars. A novel code developed in this study has been tested and verified by comparing the results of the present work with the earlier works, where ever it was possible. In particular, we performed the thermodynamic equilibrium calculations to simulate the dust condensation in circumstellar envelopes of C-rich stars. The results obtained in the present work are in good agreement with Table 2 of Lodders & Fegley (1995). This confirms the consistency of the present numerical code and the absence of any major bug or computational flaw. Hence, the thermodynamical code designed in this study is acceptable with no ambiguity in programming or discontinuity in convergence by the equation solver. Subsequent to the validation of the results of the code, we ran a set of simulations for distinct compositions of WN and WC phases of evolution of very massive stars that have been systematically studied at various dust to gas equilibrium fractions, *f*. The major features of the results have been discussed in the following.

### 4.1 Influence of non-equilibrium thermodynamics on grain condensation

Prior to substantial condensation of dust, the numerical code developed in the present work does not infer much difference among the models with distinct equilibrium criteria. All the gaseous species and the first condensate remain identical in the system irrespective of the equilibrium condition. Subsequent to the initiation of grain condensation, a fraction of the condensed dust from the system is isolated from dust to gas equilibrium. The extent of the non-equilibrium state influences the relative proportions of the condensates formed thereafter. The relative abundance of the elements is an important input component in thermodynamical calculations. The isolation of dust from equilibrium redefines the effective number density with falling temperature. The decrease in the abundance of the constituent elements modifies the system assemblage. This affects the stability fields of the subsequent condensates. An element is either removed from the system through its isolated dust, or it is locked up in its stable gaseous form. Both situations lead to a lesser number of condensates in the system. Figs. 1, 2, 5-10 (b-e) represent the mass distribution of the condensates corresponding to different values of the simulation non-equilibrium parameter, *f*. In general, the condensation sequence for non-equilibrium scenarios does not show any major distinction from the equilibrium calculations (see e.g. Tanaka, Tanaka & Nakazawa 2002). However, the more we move away from the equilibrium, the lesser is the number of condensates that are formed later in the sequence. In order to understand the condensation trends for equilibrium and non-equilibrium condensation, we have considered in details the case of non-rotating, WN and WC stars, as presented in the Figs. 1 & 2. The mass distributions of the gaseous species corresponding to the *f* value of 0.95 are shown in Figs. 3-4 for these WN and WC stars.

We assumed a completely homogenized and the vaporized environment in the initial stage of the simulation with WN composition in a non-rotating star at $10^{-3}$ bar system pressure (Figs. 1 & 3). In the equilibrium condensation calculations, TiC is the first stable condensate that appears at 1833 K in the WN phase for a non-rotating stellar model (Fig. 1a; Table 2).

**Table 2a.** Appearance (and disappearance*) temperatures (in K) of stable condensates for the WN phase in the non-rotating case at $10^{-3}$ bar pressure (WN_NR).

| Condensate | Static | Adiabatic with non-equilibrium parameter, *f* | | | |
|---|---|---|---|---|---|
| **WN_NR** | ***f* = 1** | ***f* = 1** | **0.995** | **0.99** | **0.95** |
| TiC | 1833 | 1816 | 1816 | 1816 | 1816 |
| SiC | 1691 | 1667 | 1667 | 1667 | 1667 |
| (Fe, Ni) Si | 1585 | 1542 | 1542 | 1542 | 1542 |
| SiC (*) | 1478 | 1425 | 1462 | 1492 | 1541 |
| $Fe_3C$ | 1452 | 1404 | 1415 | 1415 | - |
| (Fe, Ni) Metal | 1101 | - | - | 1368 | 1407 |
| AlN | 1421 | 1377 | 1377 | 1377 | 1377 |
| CaS | 1396 | 1350 | 1357 | 1360 | 1361 |
| $Cr_3C_2$ | 1355 | 1309 | 1297 | 1293 | 1294 |
| Co | 1279 | 1233 | 1233 | 1233 | 1233 |
| $Ni_5P_2$ | - | - | 1178 | 1199 | 1282 |
| C | 1457 | 1433 | 1171 | 1178 | 1213 |
| MgS | 1171 | 1117 | 1121 | 1125 | - |
| $Fe_3C$ (*) | 1122 | - | 1112 | - | - |
| $Si_2N_2O$ | 1121 | - | - | - | - |

**Table 2b.** Maximum value of normalized mass of stable condensates in simulated temperature range for the WN phase in the non-rotating case at $10^{-3}$ bar pressure (WN_NR).

| Condensate | Static | Adiabatic with non-equilibrium parameter, *f* |
|---|---|---|

| WN_NR | $f = 1$ | $f = 1$ | 0.995 | 0.99 | 0.95 |
|---|---|---|---|---|---|
| TiC | $8.2\times10^{-4}$ | $8.2\times10^{-4}$ | $8.2\times10^{-4}$ | $8.2\times10^{-4}$ | $8.2\times10^{-4}$ |
| SiC | $7.8\times10^{-2}$ | $8.3\times10^{-2}$ | $8.4\times10^{-2}$ | $8.4\times10^{-2}$ | $8.4\times10^{-2}$ |
| FeSi | $4.3\times10^{-1}$ | $3.9\times10^{-1}$ | $2.4\times10^{-1}$ | $1.6\times10^{-1}$ | $9.5\times10^{-2}$ |
| NiSi | $2.4\times10^{-2}$ | $2.4\times10^{-2}$ | $7.1\times10^{-2}$ | $7.0\times10^{-2}$ | $1.2\times10^{-1}$ |
| $Fe_3C$ | $1.3\times10^{-1}$ | $1.3\times10^{-1}$ | $1.9\times10^{-1}$ | $1.2\times10^{-1}$ | - |
| Fe-Metal | $2.3\times10^{-1}$ | - | - | $7.9\times10^{-2}$ | $2.2\times10^{-1}$ |
| AlN | $1.0\times10^{-2}$ | $1.0\times10^{-2}$ | $1.0\times10^{-2}$ | $1.0\times10^{-2}$ | $1.0\times10^{-2}$ |
| CaS | $2.5\times10^{-2}$ | $2.5\times10^{-2}$ | $2.5\times10^{-2}$ | $2.5\times10^{-2}$ | $2.5\times10^{-2}$ |
| $Cr_3C_2$ | $4.6\times10^{-3}$ | $4.6\times10^{-3}$ | $4.6\times10^{-3}$ | $4.6\times10^{-3}$ | $4.6\times10^{-3}$ |
| C | $4.3\times10^{-2}$ | $2.1\times10^{-2}$ | $3.8\times10^{-3}$ | $8.9\times10^{-4}$ | $9.1\times10^{-4}$ |
| MgS | $1.4\times10^{-1}$ | $1.0\times10^{-1}$ | $5.0\times10^{-2}$ | $1.0\times10^{-2}$ | - |
| $Si_2N_2O$ | $2.3\times10^{-1}$ | - | - | - | - |

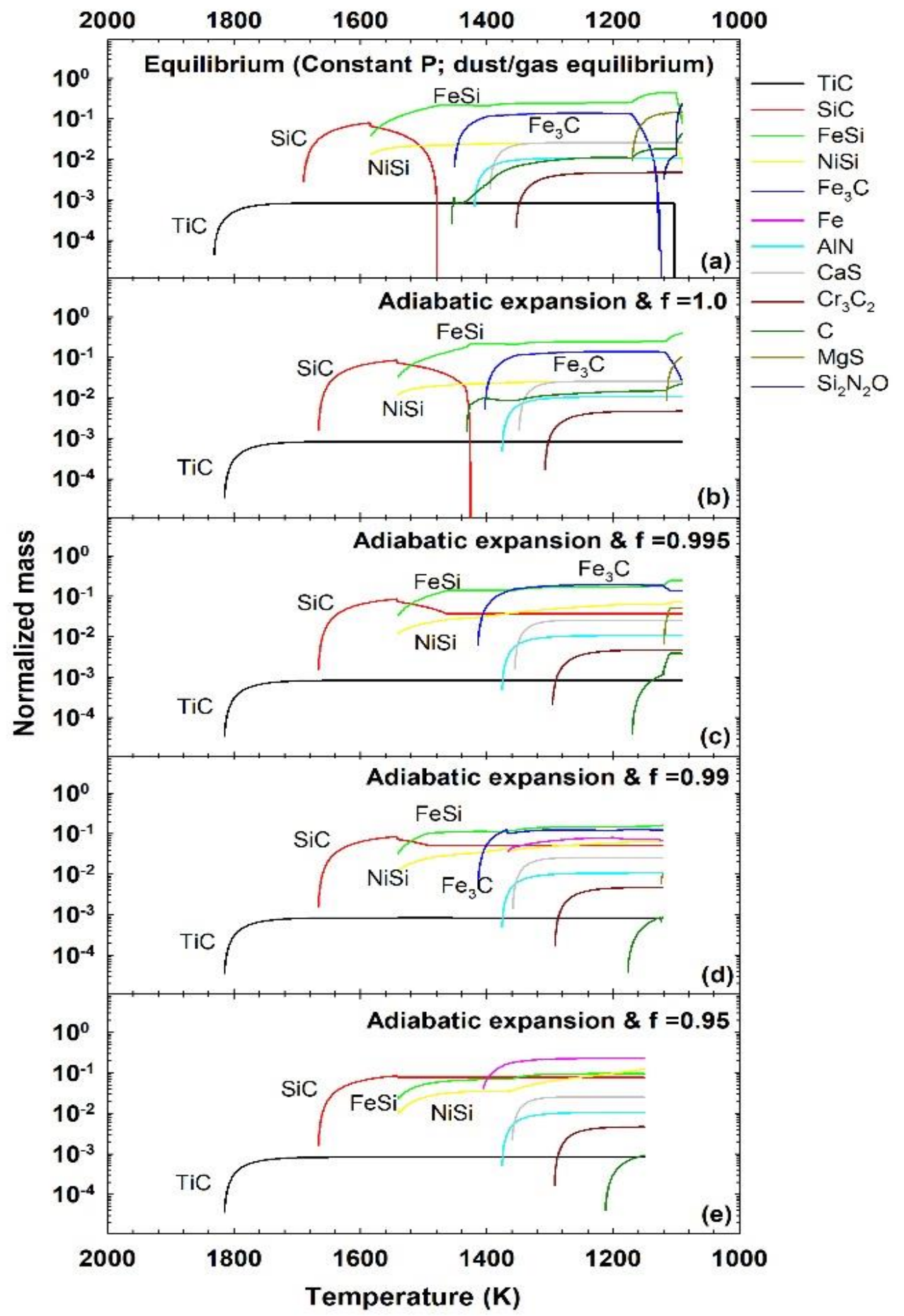

**Figure1.** The normalized mass distribution of the condensed phases as a function of temperature for a WN composition in a non-rotating stellar model at a pressure of $10^{-3}$ bar for, a) the scenario corresponding to complete dust-gas equilibrium with constant pressure. b) the scenario with dust-gas equilibrium with dynamic pressure variation in an adiabatic manner, c-e) the scenarios with the varied dust-gas non-equilibrium parameter, *f*. ranging from 0.995-0.95. In the non-equilibrium condensation scenarios, the normalized mass includes both the equilibrium as well as non-equilibrium dust components that have been removed from the chemical interaction with the residual gas. Few major species have been labelled. Other species, that are less prominent, can be seen in online colored edition.

Since the abundance of carbon is three orders of magnitude higher than titanium, the appearance of TiC does not affect the carbon species by any significant amount. The maximum amount of TiC condensation is decided by the initial abundance of titanium. The mass distribution of the species varies till 1691 K before the onset of the stability of the next condensate, SiC (beta). The condensate receives an essential amount of silicon from Si & SiH, and carbon from HCN & $C_2H_2$ gaseous species (Fig. 3). SiS is stable in gas form, and it does not decrease with the stability of solid SiC. Further, CO is the only stable gaseous form of carbon at this stage. CO forms a strong chemical bond. Its mass remains approximately the same until the low-temperature regime. Further cooling of gas below, SiC condensation, stabilizes the ferro- and nickel- silicon solid-solution. The mass of the condensate SiC in beta form increases until the appearance of the next condensate (Fe, Ni) Si at 1585 K (Fig. 1a; Table 2).

After the appearance of the new Si-bearing condensates, the mass of beta form SiC gradually decreases. As a result, SiC vanishes and completely disappears from the system assemblage at 1478 K. The availability of carbon in the system stabilizes the graphite at 1457 K. Some fraction of the released carbon reacts with the available iron gas and forms $Fe_3C$ at 1452 K. The increase in the mass of condensate and the corresponding decrease in mass of an element in the gas phase can be clearly seen with the rise and fall in their respective curves in Figs. 1(a) & 3. During the later stages, the abundance of SiS gaseous molecule decreases with the stability of CaS at 1396 K. The released fraction of silicon in turn increases (Fe, Ni) Si condensation mass. Further, silicon effectively breaks the SiS bond after the stability of MgS at 1171 K. The moment the gaseous species vanishes, the entire silicon condenses to (Fe, Ni) Si. With an increase in ferro-silicon, $Fe_3C$ vanishes at 1122 K, and, thereby, sets carbon-free to form more graphite. Finally, the carbon remains only in gaseous CO molecule form, and the solid carbon forms. $Si_2N_2O$ and TiN are the next condensates in the system. With the appearance of TiN, TiC disappears from the system. We have not shown the appearance of TiN, but the fall of TiC can clearly be seen in Fig. 1a.

In the adiabatic case of dust to gas equilibrium, the pressure decreases with the decrease in temperature. The decrease in total pressure reduces the partial pressures of the elements. Thus, the condensation temperature of all the condensates shifts downwards. The results in this scenario are almost identical to the static scenario (Fig. 1a-b; Table 2). The significant difference in this scenario is a slight upward shift in the stability of graphite, which leads to the appearance of graphite before the disappearance of SiC.

The situation is dramatically different for the other assumed values of dust to gas equilibrium parameters, *f* (Figs. 1c-e). As already explained in methodology, the non-equilibrium effects are incorporated by removing a certain fraction of dust mass from the system at every 1 K temperature fall. As mentioned earlier, the temperature drop, in the present work, is considered as a proxy for temporal evolution. Corresponding to the 0.995 value for the dust-gas equilibrium parameter, *f*, a 0.005 fraction of the dust mass is isolated from the gaseous system at every temperature step. The cumulative mass of the first condensate TiC, that includes equilibrium as well as non-equilibrium mass fractions, remains the same (Fig. 1c). Since titanium is less abundant in the system, the removal of a fraction of TiC does not produce any significant change in the system. SiC is the next condensate at 1667 K. With the removal of 0.005 fraction of SiC dust from dust-gas equilibrium with every 1 K temperature fall, the abundance of SiC almost saturates as the dust abundance, after an initial fall, is gradually stored in the form of non-equilibrium condensate that does not participate in chemical reaction (Fig. 1c). As mentioned earlier, Figures 1, 2, 5-10 present the total mass of the condensate that includes equilibrium as well as non-equilibrium contents. Thus, the constant abundance trends in any condensate, below a certain temperature, will represent substantial removal of the condensate from dust-gas equilibrium. The removal of the trapped fraction of carbon in SiC from dust-gas equilibrium at every temperature step significantly delays the stability temperature of graphite. Further decrease in the dust-gas equilibrium parameter, *f*, rapidly removes a larger fraction of condensed (SiC) dust from the system. This leads to a further reduction in the appearance of graphite (Fig. 1 c-e). The condensation of metallic iron becomes more prominent than $Fe_3C$ with a decrease in *f*. The iron condenses preferably in a metallic form (Fig. 1e) in the case of simulation with a value of 0.95 for the non-equilibrium parameter, *f*. In this scenario, a 0.05 fraction of the dust mass is isolated from the gaseous system at every temperature step. This scenario leads to the rapid removal of SiC at ~1541 K from the dust-gas equilibrium. The rapid removal of carbon favors the condensation of iron in metallic form rather than $Fe_3C$. On the appearance of CaS at 1361 K, a release of a fraction of silicon from SiS gas (Fig. 3) slightly increases the mass of (Fe, Ni) Si (Fig. 1e). The stability of MgS in the simulations with the lower non-equilibrium parameter, *f*, value is not achieved since the constituent elements get exhausted. Whenever the system is not able to supply a requisite amount of major reactive elements, the stability of condensates is not numerically feasible. In these cases, the numerical code does not converge to any solution. We stop our simulations corresponding to a temperature at which the major reactive elements get exhausted in the form of non-equilibrium dust-gas segregation. This trend is marked in all our simulations for the non-equilibrium scenarios that are stopped earlier due to the lack of numerical convergence.

The results for the distinct values of dust-gas equilibrium fractions along with the static case for the WC phase in a non-rotating star are presented in Fig. 2a-e & Fig. 4. It is clear from the normalized mass distribution of the gaseous species that CO and $C_3$ are the major gases utilizing the total carbon in the system before the temperature onset of stability of any condensate (Fig. 4). Si and $SiC_2$ utilize the maximum of silicon. The high abundance of carbon leads to the stabilization of graphite much earlier compared to other condensates at a very high temperature. After the appearance of graphite at 2783 K, all the C-containing gases except CO reduces, and the mass of the graphite increases (Fig. 2a; Table 3).

**Table 3a.** Same as Table 2 for WC phase in non-rotating case (WC_NR)

| Condensate | Static | Adiabatic with the non-equilibrium parameter, *f* | | | |
|---|---|---|---|---|---|
| **WC_NR** | ***f* = 1** | ***f* = 1** | **0.995** | **0.99** | **0.95** |
| C | 2783 | 2782 | 2782 | 2782 | 2782 |
| TiC | 1951 | 1906 | 1906 | 1906 | 1906 |
| (Fe, Ni) Si | 1487 | 1446 | 1446 | 1446 | 1446 |
| $Fe_3C$ | 1518 | 1433 | 1433 | 1432 | 1431 |
| C (*) | - | - | - | 1431 | - |
| TiC (*) | - | - | - | - | 1430 |
| (Fe, Ni) Metal | - | - | - | 1416 | - |
| $Fe_3C$ (*) | - | - | - | 1384 | - |
| CaS | 1428 | 1356 | 1357 | - | - |
| $CaAl_4O_7$ | 1420 | 1345 | 1345 | - | - |
| $Cr_3C_2$ | 1382 | 1313 | 1313 | - | - |
| Co | 1302 | 1237 | 1237 | - | - |

**Table 3b.** Same as Table 2b for WC phase in non-rotating case (WC_NR)

| Condensate | Static | Adiabatic with the non-equilibrium parameter, *f* | | | |
|---|---|---|---|---|---|
| **WC_NR** | ***f* = 1** | ***f* = 1** | **0.995** | **0.99** | **0.95** |
| C | $1.2\times10^{2}$ | $1.2\times10^{2}$ | $1.2\times10^{2}$ | $1.2\times10^{2}$ | $1.2\times10^{2}$ |
| TiC | $2.5\times10^{-3}$ | $2.5\times10^{-3}$ | $2.5\times10^{-3}$ | $2.5\times10^{-3}$ | $2.5\times10^{-3}$ |
| FeSi | $4.9\times10^{-1}$ | $5.8\times10^{-1}$ | $6.0\times10^{-1}$ | $2.0\times10^{-1}$ | $2.3\times10^{-1}$ |
| NiSi | $7.4\times10^{-2}$ | $7.4\times10^{-2}$ | $1.4\times10^{-1}$ | $5.2\times10^{-2}$ | $7.5\times10^{-2}$ |
| $Fe_3C$ | $6.5\times10^{-1}$ | $5.5\times10^{-1}$ | $5.4\times10^{-1}$ | $1.0\times10^{-1}$ | $2.6\times10^{-2}$ |
| Fe-Metal | - | - | - | $5.2\times10^{-1}$ | - |
| CaS | $7.1\times10^{-2}$ | $7.1\times10^{-2}$ | $7.2\times10^{-2}$ | - | - |
| $CaAl_4O_7$ | $1.7\times10^{-2}$ | $1.6\times10^{-2}$ | $1.6\times10^{-2}$ | - | - |
| $Cr_3C_2$ | $1.2\times10^{-2}$ | $1.3\times10^{-2}$ | $1.3\times10^{-2}$ | - | - |

This leads to a distribution of the mass of carbon either in gaseous CO or in solid graphite form. TiC is the next condensate appearing at 1951 K. In the static case (Fig. 2a), $Fe_3C$ condenses prior to (Fe, Ni) Si. This is the major distinction from the dynamic pressure scenarios having distinct dust-gas equilibrium fractions (Figs. 2b-e). In an adiabatic case, the stability field of $Fe_3C$ shifts slightly downwards, and (Fe, Ni) Si condenses prior to $Fe_3C$. This is probably due to the engagement of carbon in graphite form that is isolated due to dust-gas segregation. The rest of the condensation sequence remains the same for adiabatic cases corresponding to 1 and 0.995 values of the non-equilibrium parameter, *f*. However, the differences become prominent for the lower value of *f*. As the fraction of the condensed dust is continuously removed from the system to incorporate the non-equilibrium scenarios, the major fraction of carbon remains in gaseous CO form and reduces in condensed form (Figs. 2d-e).

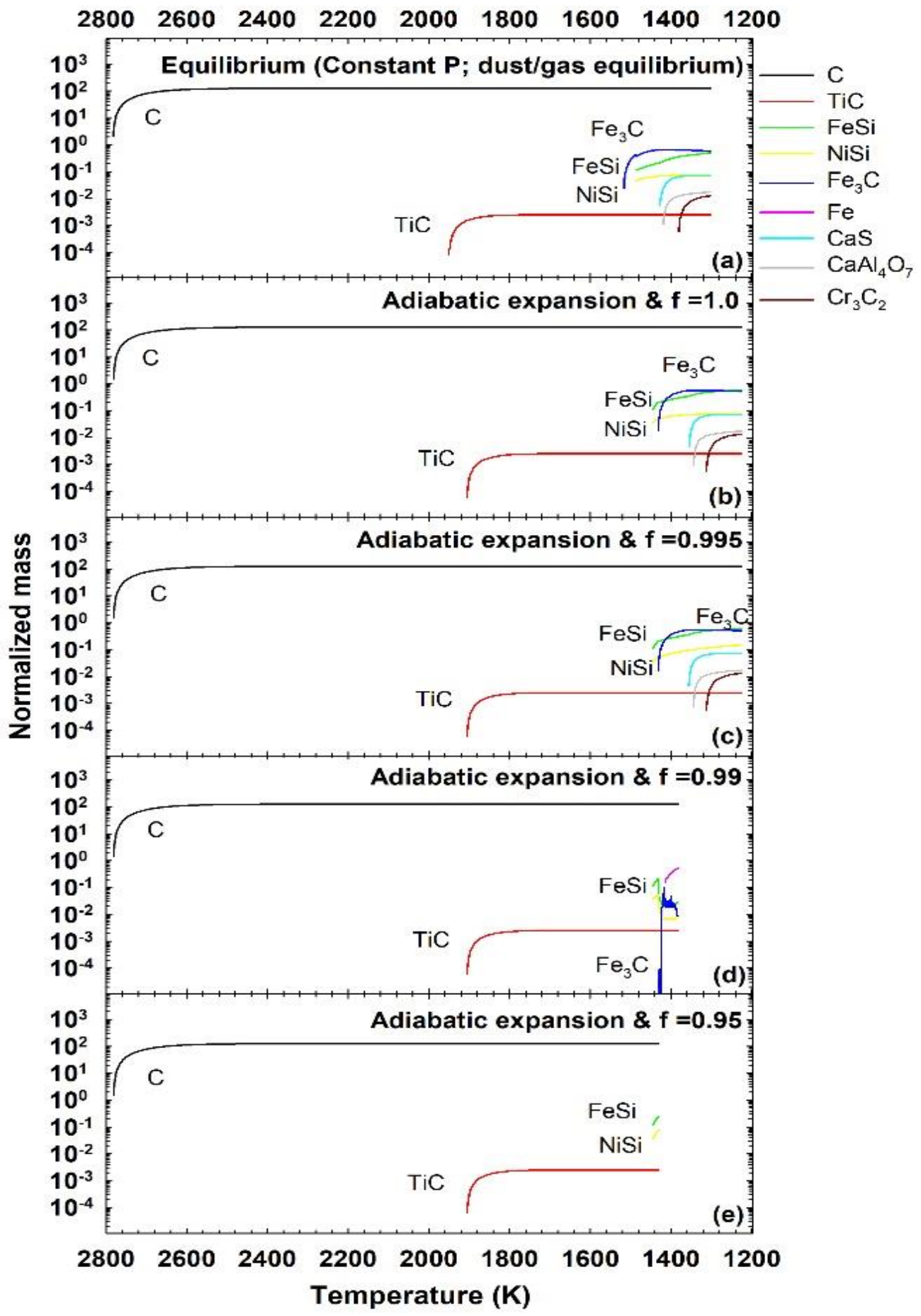

**Figure 2.** Same as Fig. 1 for WC composition in a non-rotating stellar model. The major part of the dust mass is in the carbon (graphite) form.

Thus, it affects the appearance and disappearance of further condensates. As was discussed in this subsection, the stability of TiC cannot produce any significant change in the mass distribution of carbon. So, the total dust mass remains the same even after the appearance of new condensate. The non-refractory condensates, for which the constituent elements do not exhaust, condense at the same temperature. The numerical code does not converge to yield any solution for a temperature at which the major reactive elements get exhausted. We stop our simulation at such temperatures. Compared to the WN scenario (Fig. 1), the dust mass is relatively high in WC stars (Fig. 2).

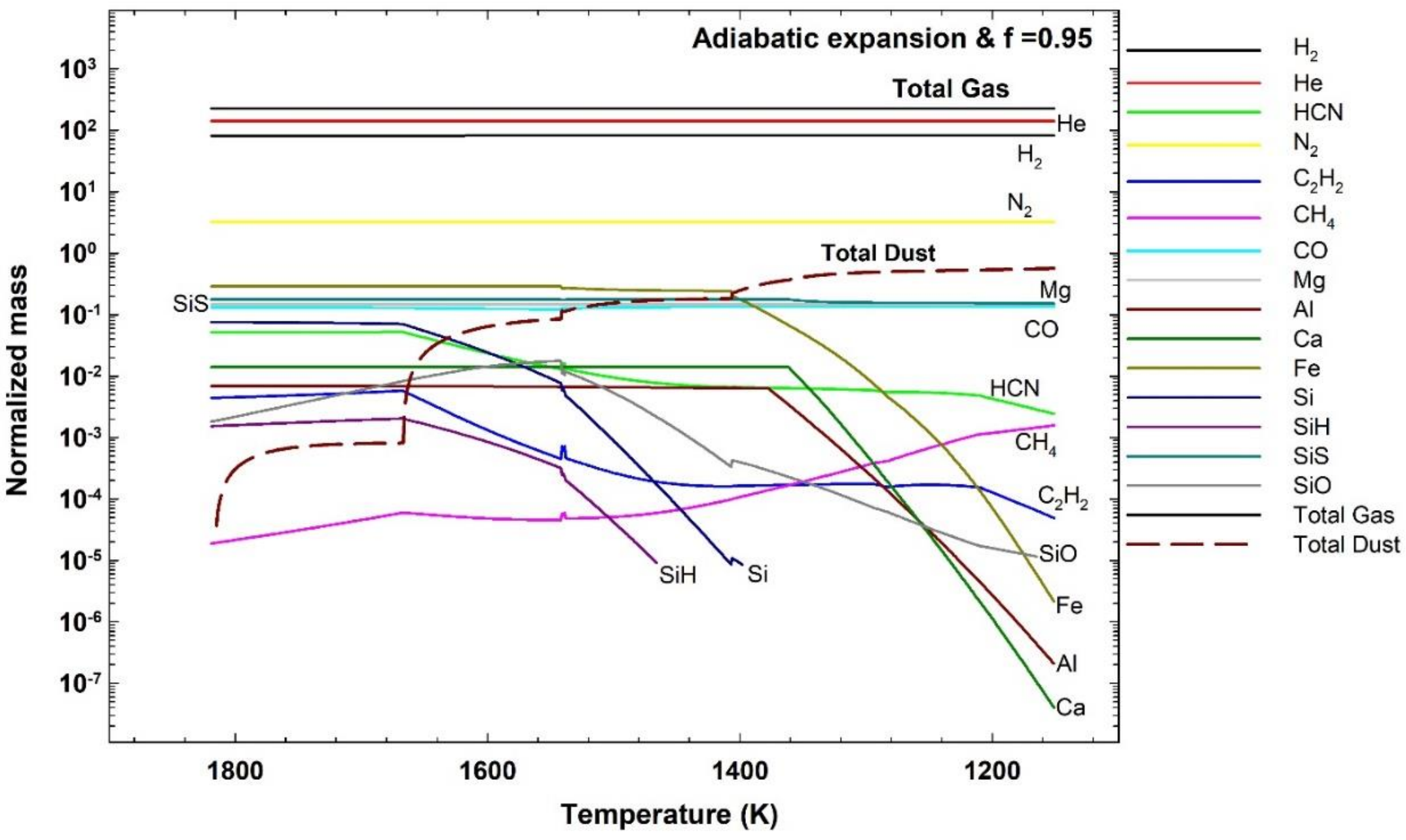

**Figure 3.** The normalized mass distribution of the gas phases as a function of temperature for a WN composition in a non-rotating stellar model at a pressure of $10^{-3}$ bar. The non-equilibrium simulation parameter, *f*, is taken to be 0.95.

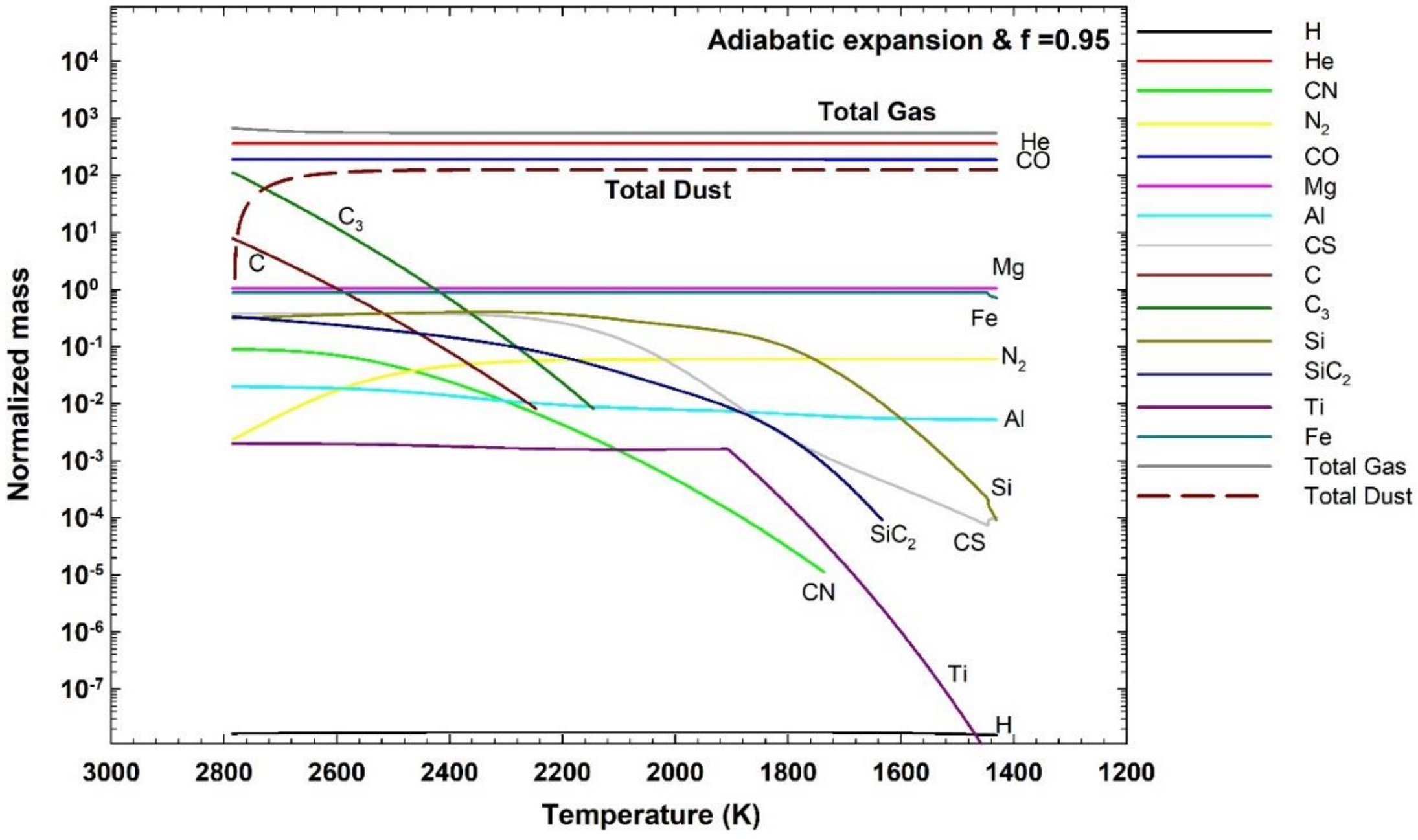

**Figure 4.** The normalized mass distribution of the gas phases as a function of temperature for a WC composition in a non-rotating stellar model at a pressure of $10^{-3}$ bar. The non-equilibrium simulation parameter, *f*, is taken to be 0.95.

## 4.2 Influence of WN and WC phases

The numerical code developed in the present work infers significant differences in the condensation sequences with the compositions corresponding to WN and WC phases. The thermodynamical condensation calculations depend upon the relative abundance of the elements in the considered system assemblage. The distinct stellar yields during WN and WC phases leads to distinct compositions in the associated circumstellar envelopes, and hence, distinct condensation sequences are produced.

The condensation temperatures for all the condensates in WN and WC compositions have been summarized in Tables 2-5 for the non-rotating and rotating stellar models.

The major features of these two stages can be compared in Tables 2 & 3 for the non-rotating stellar models, and Tables 4 & 5 for the rotating stellar models. The normalized mass distributions of condensed phases for WN and WC compositions in non-rotating and rotating stellar models corresponding to $10^{-3}$ bar pressure are shown in Figs. 1, 2 and 5, 6, respectively.

**Table 4a.** Same as Table 2 for WN phase in rotating case (WN_R)

| Condensate | Static | Adiabatic with the non-equilibrium parameter, *f* | | | |
|---|---|---|---|---|---|
| WNR | *f* = 1 | *f* = 1 | 0.995 | 0.99 | 0.95 |
| TiN | 1642 | 1602 | 1602 | 1602 | 1602 |
| $CaAl_4O_7$ | 1572 | 1547 | 1547 | 1547 | 1547 |
| CaS | 1486 | 1453 | 1453 | 1453 | 1453 |
| (Fe, Ni) Metal | 1462 | 1415 | 1415 | 1415 | 1415 |
| $Ni_3P$ | - | - | - | - | 1336 |
| Melilite | 1351 | 1317 | 1317 | 1317 | 1317 |
| (Fe, Ni) Si | 1338 | 1289 | 1288 | - | - |
| $CaAl_4O_7$ (*) | 1330 | 1282 | 1287 | 1305 | 1316 |
| $Cr_5Si_3$ | 1326 | 1278 | 1277 | 1277 | - |
| CaS (*) | 1289 | - | - | - | - |
| Co | 1279 | - | - | - | - |
| Spinel | 1273 | - | - | - | - |
| $CaMgSiO_4$ | 1269 | - | - | - | - |
| Melilite (*) | 1261 | - | - | - | - |
| Olivine | 1257 | - | - | - | - |

**Table 4b.** Same as Table 2b for WN phase in rotating case (WN_R)

| Condensate | Static | Adiabatic with the non-equilibrium parameter, *f* | | | |
|---|---|---|---|---|---|
| WNR | *f* = 1 | *f* = 1 | 0.995 | 0.99 | 0.95 |
| TiN | $8.5\times10^{-4}$ | $8.5\times10^{-4}$ | $8.5\times10^{-4}$ | $8.5\times10^{-4}$ | $8.5\times10^{-4}$ |
| $CaAl_4O_7$ | $1.6\times10^{-2}$ | $1.6\times10^{-2}$ | $1.6\times10^{-2}$ | $1.6\times10^{-2}$ | $1.6\times10^{-2}$ |
| CaS | $1.9\times10^{-2}$ | $1.9\times10^{-2}$ | $1.9\times10^{-2}$ | $1.9\times10^{-2}$ | $2.0\times10^{-2}$ |
| Fe-Metal | $2.7\times10^{-1}$ | $2.7\times10^{-1}$ | $2.8\times10^{-1}$ | $2.8\times10^{-1}$ | $2.8\times10^{-1}$ |
| $Ca_2Al_2SiO_7$ | $3.4\times10^{-2}$ | $3.4\times10^{-2}$ | $1.2\times10^{-2}$ | $4.9\times10^{-3}$ | $1.0\times10^{-4}$ |
| $Ca_2MgSi_2O_7$ | $1.9\times10^{-2}$ | $5.3\times10^{-3}$ | $5.3\times10^{-3}$ | $1.2\times10^{-3}$ | $1.0\times10^{-6}$ |
| FeSi | $1.5\times10^{-1}$ | $1.1\times10^{-2}$ | $5.3\times10^{-3}$ | - | - |
| NiSi | $1.2\times10^{-2}$ | $1.3\times10^{-3}$ | $1.1\times10^{-3}$ | - | - |
| $Cr_5Si_3$ | $5.2\times10^{-3}$ | $2.8\times10^{-3}$ | $2.8\times10^{-3}$ | $3.1\times10^{-3}$ | - |
| $MgAl_2O_4$ | $1.8\times10^{-2}$ | - | - | - | - |
| $FeAl_2O_4$ | $5.2\times10^{-7}$ | - | - | - | - |
| $CaMgSiO_4$ | $5.4\times10^{-2}$ | - | - | - | - |

**Table 5a.** Same as Table 2 but for the WC phase in rotating case (WC_R)

| Condensate | Static | Adiabatic with equilibrium fraction, f | | | |
|---|---|---|---|---|---|
| **WCR** | ***f* = 1** | ***f* = 1** | **0.995** | **0.99** | **0.95** |
| C | 2743 | 2739 | 2739 | 2739 | 2739 |
| TiC | 1949 | 1903 | 1903 | 1903 | 1903 |
| SiC | 1673 | 1643 | 1643 | 1643 | 1643 |
| C (*) | - | - | - | 1603 | 1637 |
| (Fe, Ni) Si | 1520 | 1446 | 1446 | 1461 | 1464 |
| SiC (*) | 1516 | 1435 | 1444 | 1457 | 1462 |
| $Fe_3C$ | 1487 | 1407 | 1416 | - | - |
| (Fe, Ni) Metal | - | - | - | 1407 | - |
| CaS | 1416 | 1346 | 1348 | 1352 | - |
| AlN | 1397 | 1330 | 1330 | - | - |
| $CaAl_4O_7$ | 1379 | 1290 | 1288 | 1369 | - |
| $Cr_5Si_3$ | - | - | - | 1273 | - |
| $Cr_3C_2$ | 1373 | 1305 | 1305 | - | - |
| AlN (*) | 1376 | 1288 | 1286 | - | - |

**Table 5b.** Same as Table 2b but for the WC phase in rotating case (WC_R)

| Condensate | Static | Adiabatic with equilibrium fraction, f | | | |
|---|---|---|---|---|---|
| **WCR** | ***f* = 1** | ***f* = 1** | **0.995** | **0.99** | **0.95** |
| C | $6.9\times10^{1}$ | $6.9\times10^{1}$ | $6.9\times10^{1}$ | $6.9\times10^{1}$ | $7.0\times10^{1}$ |
| TiC | $1.8\times10^{-3}$ | $1.8\times10^{-3}$ | $1.8\times10^{-3}$ | $1.8\times10^{-3}$ | $1.8\times10^{-3}$ |
| SiC | $1.1\times10^{-1}$ | $1.5\times10^{-1}$ | $1.5\times10^{-1}$ | $7.0\times10^{-2}$ | $2.8\times10^{-2}$ |
| FeSi | $4.7\times10^{-1}$ | $4.8\times10^{-1}$ | $3.4\times10^{-1}$ | $1.5\times10^{-1}$ | $1.1\times10^{-1}$ |
| NiSi | $5.4\times10^{-2}$ | $5.4\times10^{-2}$ | $9.0\times10^{-2}$ | $4.4\times10^{-2}$ | $4.7\times10^{-2}$ |
| $Fe_3C$ | $3.6\times10^{-1}$ | $3.3\times10^{-1}$ | $4.4\times10^{-1}$ | - | - |
| Fe-Metal | - | - | - | $5.2\times10^{-1}$ | - |
| CaS | $4.8\times10^{-2}$ | $5.2\times10^{-2}$ | $5.3\times10^{-2}$ | $4.8\times10^{-2}$ | - |
| AlN | $7.4\times10^{-3}$ | $1.1\times10^{-2}$ | $1.2\times10^{-2}$ | - | - |
| $Cr_3C_2$ | $9.7\times10^{-3}$ | $8.0\times10^{-3}$ | $4.2\times10^{-3}$ | - | - |
| $Cr_5Si_3$ | - | - | - | $5.4\times10^{-3}$ | - |
| $CaAl_4O_7$ | $2.5\times10^{-2}$ | $2.3\times10^{-2}$ | $1.7\times10^{-2}$ | $2.6\times10^{-2}$ | - |

The four most abundant elements in the WN composition for the non-rotating stellar model are H, He, N and C. The abundance of C relative to O is the most prominent ratio which affects the gas grain chemistry the most. Carbon and oxygen react to form CO molecule which is very stable at high temperatures. The less abundant element out of the two is consumed fully in this strong chemical bond, whereas, the remaining fraction of the more abundant element can form other species. Since the C/O ratio is ~1.48 in the WN composition, ~67 per cent of the C is locked up in the CO molecule. Other dominating C-bearing gaseous species are HCN, $C_2H_2$, and $CH_4$ (Fig. 3). In the case of WC composition for a non-rotating stellar model, He, C, O, and Mg are the four most abundant elements. The C/O ratio is ~2.53 in the WC composition, and CO, $C_3$ are the major gases utilizing the total carbon in the system before the onset of stability of any condensate (Fig. 4). The major difference between the compositions of WN and WC phases in non-rotating stellar model, which affect the condensation sequence, are the abundances of N, C and O. Although the abundance of carbon is more than the abundance of oxygen in both the compositions, the condensation sequence of C, TiC and SiC varies according to the value of C/O ratio. TiC, SiC, and C is the order of appearance in the case of WN composition (Fig. 1a; Table 2). The stable gases like HCN and CN are formed, via, the thermochemical reaction of CO or $C_2H_2$ with $N_2$ and $H_2$ molecules due to an increase in the N abundance. This reduces the condensation temperature of graphite substantially, and the other C-bearing condensates like carbides slightly. However, nitrides cannot replace the carbides as the initial condensates in the case of the WN phase. On the other hand, the high abundance of carbon allows the condensation of graphite much prior to TiC and SiC in C-rich envelopes of WC star for which C/O ratio is greater than 2 (Fig. 2a; Table 3). The result is in good agreement with Table 3 of Lodders and Fegley (1995).

## 4.3 Influence of stellar rotation on grain condensation

The composition of the system assemblage is an essential input component in thermodynamical condensation calculations. As already explained in the introduction section, the rotation affects the evolutionary track of a star. It has a significant effect on all the stellar outputs. Thus, the non-rotating and rotating stellar models have distinct stellar yields, which lead to distinct relative abundances of the elements in the circumstellar envelopes of non-rotating and rotating stars. Hence, the condensation chemistry of the elements gets modified accordingly and produces distinct sequences in the condensation calculations. The condensation temperatures for all the condensates for the rotating as well as non-rotating stellar models have been summarized in Tables 2-5 in WN and WC phases. The normalized mass distribution of condensed phases for WN and WC compositions have been shown in Figs. 1, 2, 5 & 6. The influence of rotation on the condensation of dust grains can be compared in Tables 2 & 4 and Figs. 1 & 5 for the WN phase, and Tables 3 & 5 and Figs. 2 & 6 for WC phase. The major features of the results of WN and WC phases for rotating stellar models in the dust-gas equilibrium scenario with constant pressure have been presented in Figs. 5a and 6a. Only the rotating stellar models with constant total pressure are discussed in this subsection, as the non-rotating stellar models have been already discussed.

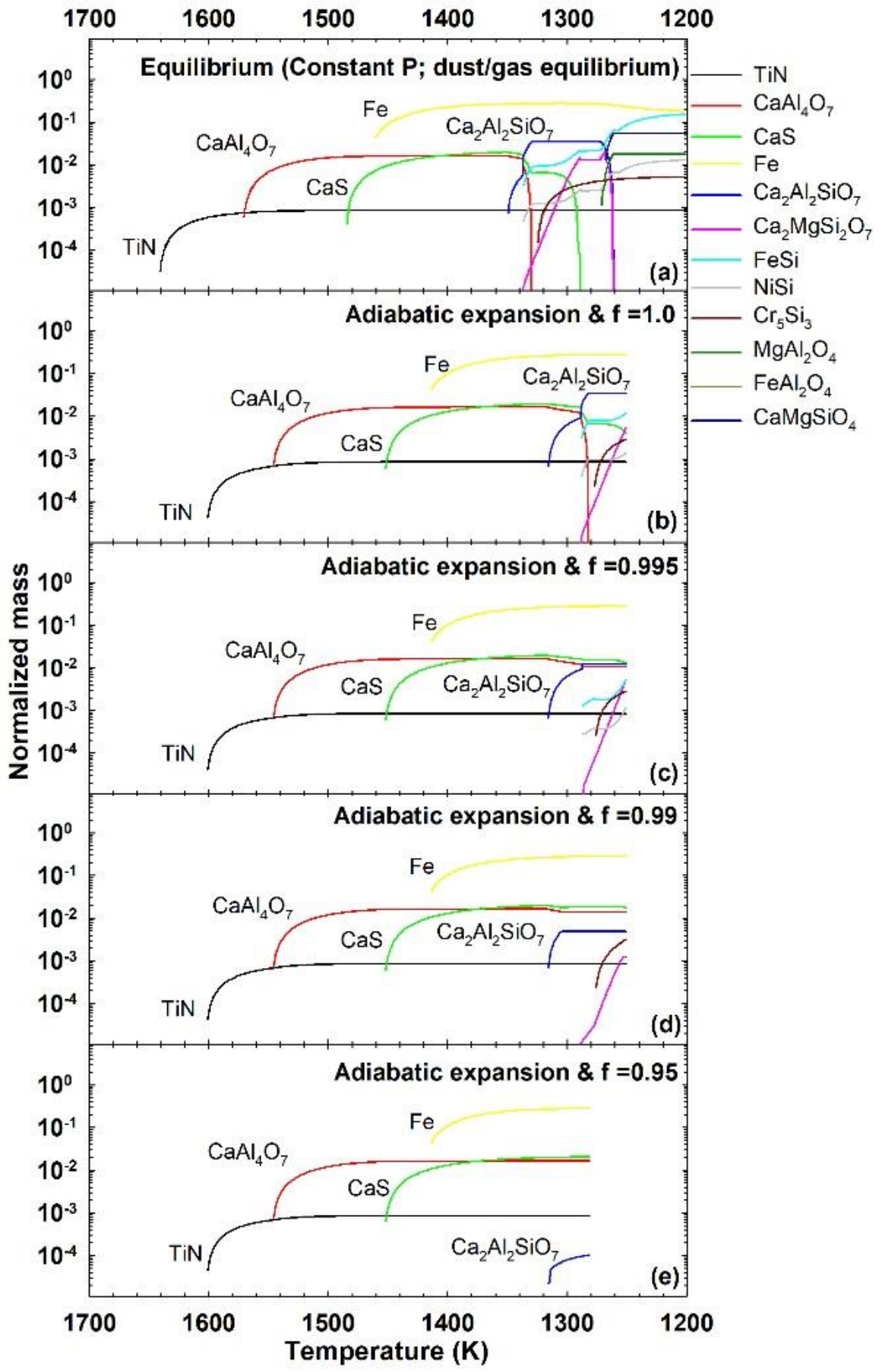

**Figure 5**. Same as Fig. 1 for WN composition in a rotating stellar model. The major part of the dust mass is in the metallic iron form.

In the case of WN phase, the abundance of oxygen increases for rotating stellar model which makes oxygen fourth in terms of abundance after H, He and N. This increase in oxygen abundance leads to C/O ratio below 1 which is more than 1 for a non-rotating stellar model in WN phase. The less abundant carbon is consumed fully in the strong CO bond, whereas, the remaining fraction of the more abundant oxygen can form O-bearing species. Therefore, no carbides are formed in the rotating stellar model and only the oxides and silicates get stabilized unlike the non-rotating stellar model of the WN phase (Figs. 1a and 5a). The abundance of carbon also affects the formation of nitrides. In the non-rotating stellar model, TiC stabilizes first which delays the appearance of TiN, whereas the rotating model does not allow the formation of carbides and thus, TiN condenses as the first condensate. Subsequent condensates include iron in metal form or O- and Si- bearing condensates like grossite, melilite, olivine, etc. (Fig. 5a; Table 4).

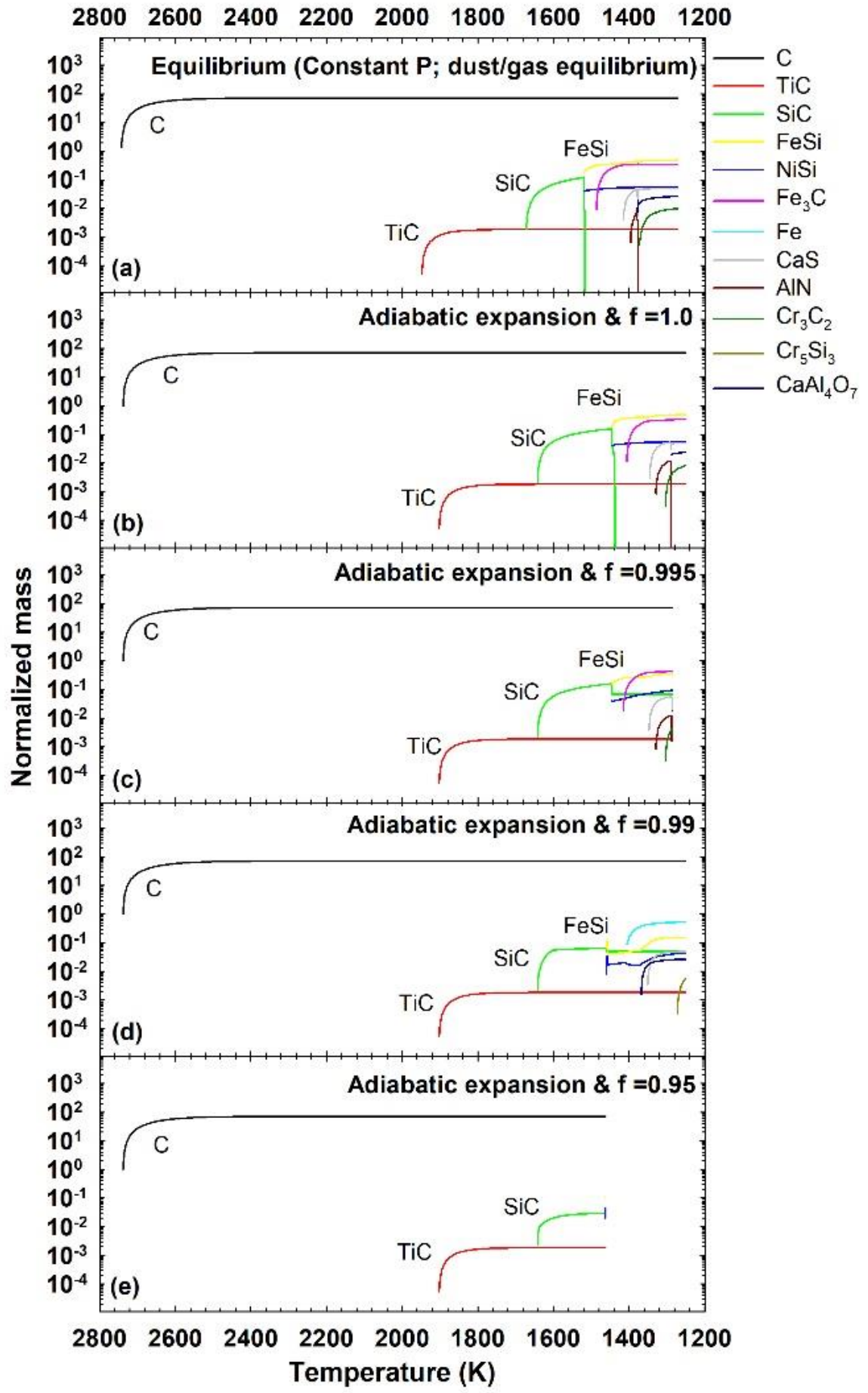

**Figure 6.** Same as Fig. 1 for WC composition in a rotating stellar model. The major part of the dust mass is in the carbon (graphite) form.

Similarly, in case of WC phase, the abundance of nitrogen increases for rotating stellar model which makes nitrogen fourth in terms of abundance after He, C and O. An increase in the abundance of nitrogen allowed the condensation of AlN in the rotating stellar model, which does not appear in non-rotating stellar model in WC phase. Further, the abundance of oxygen decreases which causes the increase in the C/O ratio from 2.5 in a non-rotating model to ~4.9 in the rotating stellar model. As already explained in previous subsections, a high abundance of carbon relative to the abundance of oxygen increases the availability of carbon in the system which, in turn, affects the condensation of carbides and oxides. Thus, the condensation temperature of SiC shifts upwards, and the condensation temperature of oxides like grossite shifts downwards from 1420 K to 1379 K (Figs. 2a and 6a). The major difference between the features of rotating and non-rotating stellar models in the WC phase is the appearance and disappearance of SiC. The other features like the formation of graphite and carbides at a higher temperature and much earlier to O-bearing condensates are similar to its non-rotating counterpart (Fig. 6a; Table 5).

## 4.4 Influence of late and early phases of WN

We have also explored the condensation sequence for the late and early-type WN stars. The condensation of dust grains depends upon the composition of the system assemblage. As was explained in the introduction section, the early and late-type are the subclasses of WN stars based on high or low ionization. WNE is the evolutionary stage of the WR phase which appears after the WNL stage. The H-emission line is shown by the WNL phase but is absent in the WNE phase. The chemical compositions are different in these two phases that affect the type of dust which can stabilize in that environment, and hence produces distinct condensation sequences. The names of all the condensates and their corresponding condensation temperatures for WNL and WNE compositions are presented in Tables 6-7. The normalized mass distributions of the condensed phases for WNL and WNE compositions corresponding to $10^{-3}$ bar pressure at different dust to gas equilibrium fractions are shown in Figs. 7, 8. The major features of the results of WNL and WNE phases in the dust-gas equilibrium scenario with constant pressure are presented in Figs. 7a, 8a.

**Table 6a.** Same as Table 2 for late-type WN phase (WNL)

| Condensate | Static | Adiabatic with the non-equilibrium parameter, *f* | | | |
|---|---|---|---|---|---|
| **WNL** | ***f* = 1** | ***f* = 1** | **0.995** | **0.99** | **0.95** |
| TiC | 1805 | 1785 | 1785 | 1772 | 1785 |
| SiC | 1671 | 1643 | 1643 | 1631 | 1643 |
| (Fe, Ni) Si | 1463 | 1418 | 1418 | 1405 | 1418 |
| AlN | 1427 | 1382 | 1382 | 1370 | 1382 |
| $Cr_5Si_3$ | 1418 | 1373 | 1373 | 1361 | 1373 |
| Co | 1393 | 1347 | 1347 | 1335 | 1347 |
| TiN | 1355 | 1303 | 1303 | 1290 | 1334 |
| TiC (*) | 1355 | 1303 | 1303 | 1290 | 1334 |
| CaS | 1334 | 1284 | 1284 | 1272 | - |
| $Si_2N_2O$ | 1307 | 1258 | 1258 | 1247 | - |
| $CaAl_4O_7$ | - | - | - | 1203 | - |
| Spinel | - | - | 1116 | - | - |
| MgS | 1156 | 1101 | 1111 | - | - |
| Spinel (*) | - | - | 1108 | - | - |
| $Cr_3C_2$ | 1147 | 1095 | - | - | - |
| $Cr_5Si_3$ (*) | 1144 | 1094 | - | - | - |
| C | 1133 | 1080 | - | - | - |
| SiC (*) | 1125 | 1071 | 1189 | 1232 | 1368 |
| $Ni_5P_2$ | 1107 | 1062 | - | - | - |
| $Fe_3C$ | 1094 | 1041 | - | - | - |
| (Fe, Ni) Si (*) | 1093 | 1040 | - | - | - |
| MgO | 1071 | 1015 | - | - | - |

**Table 6b.** Same as Table 2b for late-type WN phase (WNL)

| Condensate | Static | Adiabatic with the non-equilibrium parameter, *f* | | | |
|---|---|---|---|---|---|
| **WNL** | ***f* = 1** | ***f* = 1** | **0.995** | **0.99** | **0.95** |
| TiC | $1.1\times10^{-3}$ | $1.1\times10^{-3}$ | $1.1\times10^{-3}$ | $1.0\times10^{-3}$ | $1.1\times10^{-3}$ |
| SiC | $9.6\times10^{-2}$ | $1.0\times10^{-1}$ | $9.2\times10^{-2}$ | $8.1\times10^{-2}$ | $8.5\times10^{-2}$ |
| NiSi | $1.5\times10^{-3}$ | $1.5\times10^{-3}$ | $1.5\times10^{-3}$ | $1.3\times10^{-3}$ | $1.5\times10^{-3}$ |
| FeSi | $6.3\times10^{-3}$ | $6.3\times10^{-3}$ | $1.4\times10^{-2}$ | $1.4\times10^{-2}$ | $1.3\times10^{-2}$ |
| AlN | $1.4\times10^{-2}$ | $1.4\times10^{-2}$ | $1.4\times10^{-2}$ | $1.2\times10^{-2}$ | $1.1\times10^{-2}$ |
| $Cr_5Si_3$ | $7.3\times10^{-3}$ | $7.3\times10^{-3}$ | $7.3\times10^{-3}$ | $6.3\times10^{-3}$ | $4.2\times10^{-3}$ |
| TiN | $1.1\times10^{-3}$ | $1.1\times10^{-3}$ | $1.2\times10^{-4}$ | $1.1\times10^{-5}$ | - |
| CaS | $3.4\times10^{-2}$ | $3.5\times10^{-2}$ | $3.5\times10^{-2}$ | $3.0\times10^{-2}$ | - |
| $Si_2N_2O$ | $3.9\times10^{-1}$ | $3.9\times10^{-1}$ | $1.4\times10^{-1}$ | $6.4\times10^{-2}$ | - |
| $CaAl_4O_7$ | - | - | - | $4.0\times10^{-3}$ | - |
| $MgAl_2O_4$ | - | - | $1.0\times10^{-4}$ | - | - |
| $FeAl_2O_4$ | - | - | $2.2\times10^{-11}$ | - | - |
| MgS | $2.0\times10^{-1}$ | $2.0\times10^{-1}$ | $3.8\times10^{-2}$ | - | - |
| $Cr_3C_2$ | $6.4\times10^{-3}$ | $6.4\times10^{-3}$ | - | - | - |
| C | $6.6\times10^{-2}$ | $6.4\times10^{-2}$ | - | - | - |

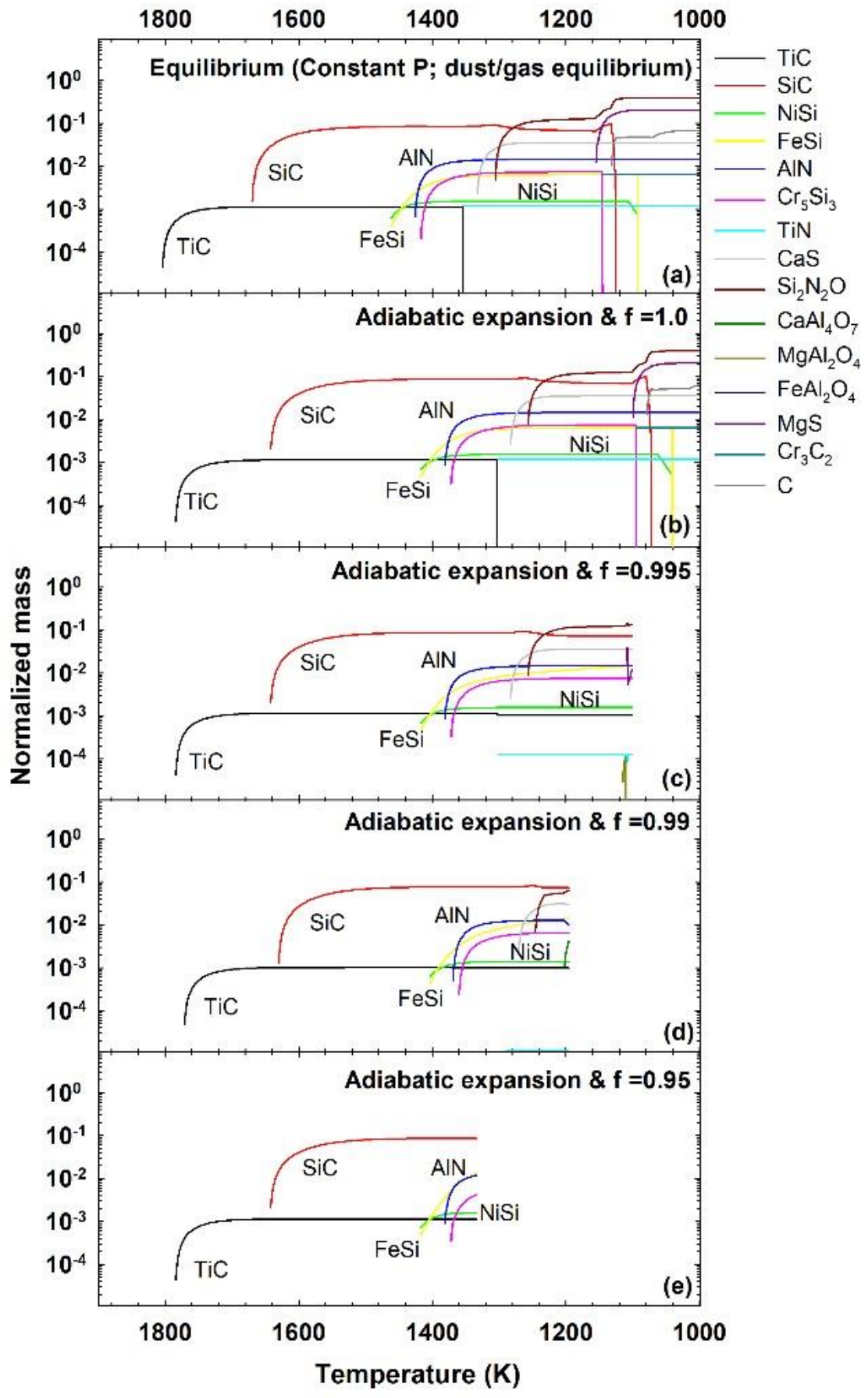

**Figure 7**. Same as Fig. 1 for the late-type WN phase. The major part of the dust mass is in carbide or nitride form.

**Table 7a.** Same as Table 2 for early-type WN phase (WNE)

| Condensate | Static | Adiabatic with the non-equilibrium parameter, *f* | | | |
|---|---|---|---|---|---|
| **WNE** | ***f* = 1** | ***f* = 1** | **0.995** | **0.99** | **0.95** |
| C | 2221 | 2206 | 2206 | 2206 | 2206 |
| TiC | 1943 | 1910 | 1910 | 1910 | 1910 |
| SiC | 1776 | 1730 | 1730 | 1730 | 1730 |
| C (*) | - | - | 1712 | 1726 | 1728 |
| (Fe, Ni) Si | 1355 | 1294 | 1417 | 1426 | 1428 |
| AlN | 1434 | 1377 | 1377 | 1377 | 1377.3 |
| $Cr_5Si_3$ | - | - | 1370 | 1375 | 1377.4 |
| Co | 1400 | 1341 | 1341 | 1341 | - |
| TiN | - | - | 1307 | 1325 | - |
| TiC (*) | - | - | 1307 | 1325 | - |
| CaS | 1398 | 1338 | 1277 | 1272 | - |
| $Si_2N_2O$ | - | - | 1252 | 1260 | - |
| SiC (*) | - | - | - | 1247 | 1376 |
| $Cr_3C_2$ | 1367 | 1309 | 1081 | - | - |
| MgS | 1169 | 1102 | 1094 | - | - |

**Table 7b.** Same as Table 2b for early-type WN phase (WNE)

| Condensate | Static | Adiabatic with the non-equilibrium parameter, *f* | | | |
|---|---|---|---|---|---|
| **WNE** | ***f* = 1** | ***f* = 1** | **0.995** | **0.99** | **0.95** |
| C | $1.0\times10^{-1}$ | $1.0\times10^{-1}$ | $1.0\times10^{-1}$ | $1.0\times10^{-1}$ | $1.0\times10^{-1}$ |
| TiC | $1.5\times10^{-3}$ | $1.5\times10^{-3}$ | $1.5\times10^{-3}$ | $1.5\times10^{-3}$ | $1.5\times10^{-3}$ |
| SiC | $3.7\times10^{-1}$ | $3.6\times10^{-1}$ | $1.5\times10^{-1}$ | $1.1\times10^{-1}$ | $1.0\times10^{-1}$ |
| NiSi | $2.1\times10^{-3}$ | $2.1\times10^{-3}$ | $2.1\times10^{-3}$ | $2.1\times10^{-3}$ | $2.1\times10^{-3}$ |
| FeSi | $8.8\times10^{-3}$ | $8.8\times10^{-3}$ | $2.1\times10^{-2}$ | $2.2\times10^{-2}$ | $1.0\times10^{-2}$ |
| AlN | $2.0\times10^{-2}$ | $2.0\times10^{-2}$ | $2.0\times10^{-2}$ | $1.9\times10^{-2}$ | $6.8\times10^{-3}$ |
| $Cr_5Si_3$ | - | - | $1.0\times10^{-2}$ | $9.9\times10^{-3}$ | $1.4\times10^{-3}$ |
| TiN | - | - | $1.0\times10^{-4}$ | $8.8\times10^{-6}$ | - |
| CaS | $4.8\times10^{-2}$ | $4.8\times10^{-2}$ | $4.8\times10^{-2}$ | $4.8\times10^{-2}$ | - |
| $Si_2N_2O$ | - | - | $2.3\times10^{-1}$ | $8.0\times10^{-2}$ | - |
| MgS | $2.0\times10^{-1}$ | $1.7\times10^{-1}$ | $1.5\times10^{-1}$ | - | - |
| $Cr_3C_2$ | $8.9\times10^{-3}$ | $8.9\times10^{-3}$ | - | - | - |

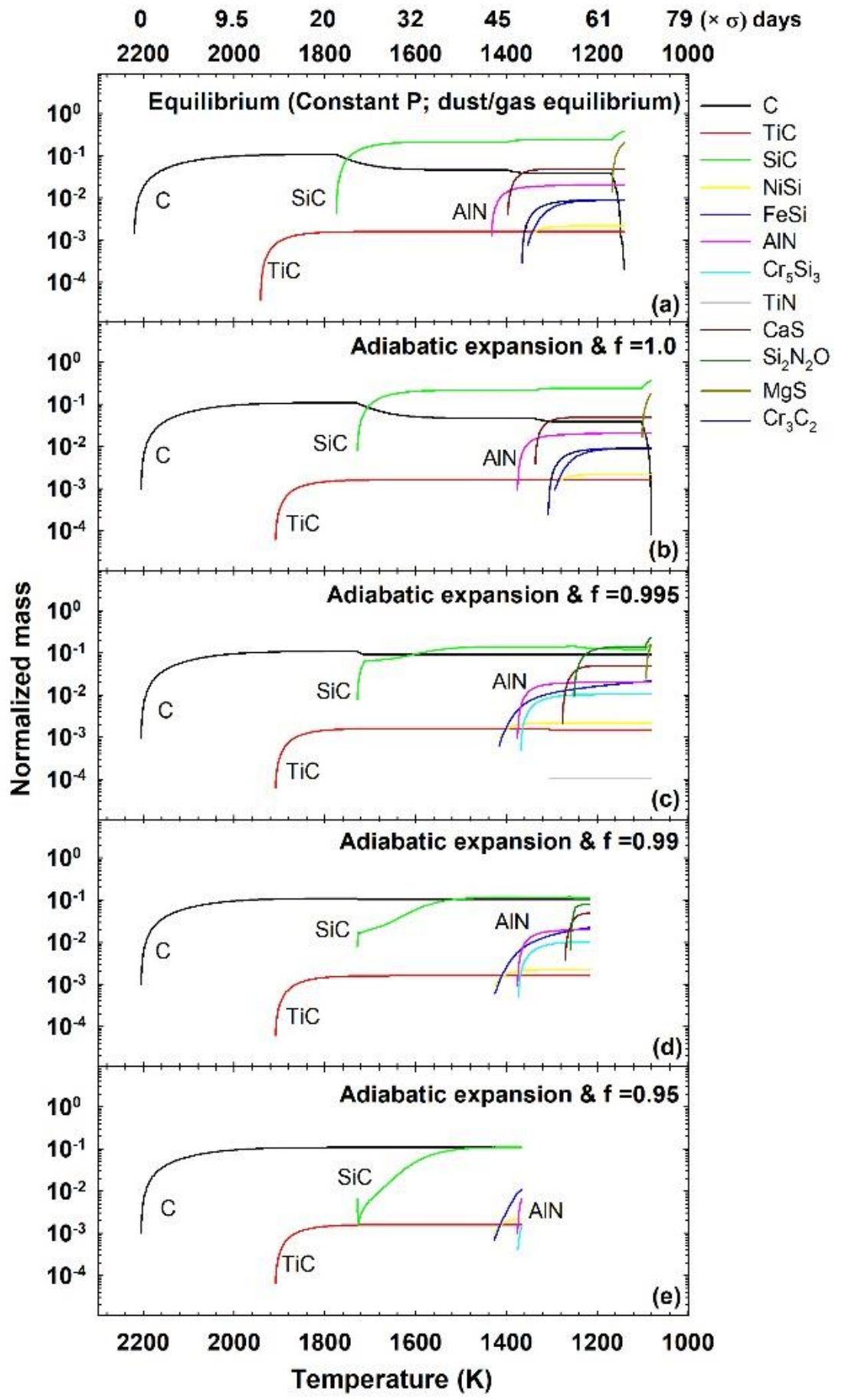

**Figure 8.** Same as Fig. 1 for the early-type WN phase. The major part of the dust mass is in carbon (graphite) or in carbide form. A possible time-scale associated with condensation of a dust grain is presented in the topmost panel. The grain growth has been considered spherically initiating from a 1 $\AA$ size to a final radius of 1 µm. Here, σ is a multiple that can provide condensation over a time-scale of tens to hundreds of days.

N, Mg, and Si are the abundant elements after He and H in the case of WNL composition. Nitrides like AlN, TiN, $Si_2N_2O$ are formed in the WNL phase (Fig. 7a: Table 6). On the contrary, carbon is abundant after, He and N in the case of the WNE phase which favours the formation of graphite and carbides predominantly. Refractory nitride AlN is formed only (Fig. 8a: Table 7). The C/O ratio is more than unity in both the phases, which drops the condensation temperatures of oxides and silicates significantly and supports the formation of carbides, nitrides, and sulphides. TiC, SiC, and C is the sequence of condensation in case of WNL composition, whereas C, TiC and SiC is the order of appearance in WNE composition (Figs. 7a, 8a). As already discussed in subsection 4.1, the distinction is primarily caused by C/O ratio which is closer to WN star in a non-rotating model for the WNL phase, and WC star in a non-rotating model for WNE phase. The major features of the WNL and WNE phases match quite well with the features of WN and WC stars respectively due to their similar compositions.

## 4.5 Influence of variation in pressure on grain condensation

The condensation temperature of dust grains depends directly on the total pressure of the system assemblage. The variation in the total pressure results in the change of partial pressure of the elements, and thus, the stability field of the condensates. At high partial pressures, the condensation temperatures can exceed the melting points of the individual species. This results in condensation at higher temperatures. In the present study, we have explored six distinct compositions of WR stars which include the WN and WC phases for non-rotating and rotating stellar models, apart from the WN star in late and early phases. The variation in pressure has been explored in two distinct scenarios. In the first scenario, the total pressure of the system assemblage has been assigned a dynamical value for the simulations and allowed to decrease adiabatically with the cooling of the system. The shift in condensation temperatures with the adiabatic variation of pressure can be seen by comparing the graphs in the panels (a) and (b) of the figures representing all the considered models (Figs. 1, 2, 5-10; Tables 2-9). In the second scenario, a distinct initial total pressure of the system has been chosen. Two distinct sets of simulations have been performed for the WC phase in the non-rotating stellar model and late-type WN phase with an initial total pressure of $10^{-6}$ bar. The temperatures of appearance and disappearance of all the condensates for these two simulations have been summarized in Tables 8 & 9. The corresponding mass distributions of condensed phases are shown in Fig. 9 &10. The influence of using a distinct value of system total pressure can be seen by comparing Figs. 2 & 9, and Fig. 7 & 10.

**Table 8a.** Same as Table 2 for WC phase in the non-rotating case at $10^{-6}$ bar pressure (WC_NR_P6)

| Condensate | Static | Adiabatic with the non-equilibrium parameter, *f* | | | |
|---|---|---|---|---|---|
| **WC_NR_P6** | ***f* = 1** | ***f* = 1** | **0.995** | **0.99** | **0.95** |
| C | 2313 | 2301 | 2301 | 2301 | 2301 |
| TiC | 1666 | 1629 | 1629 | 1629 | 1629 |
| SiC | 1390 | 1375 | 1375 | 1375 | 1375 |
| C (*) | - | - | - | 1326 | 1371 |
| (Fe, Ni) Si | 1261 | 1206 | 1206 | 1216 | 1220 |
| SiC (*) | 1257 | 1197 | 1204 | - | - |
| $Fe_3C$ | 1226 | 1167 | 1174 | - | - |
| CaS | 1190 | 1136 | 1138 | - | - |
| $CaAl_4O_7$ | 1172 | 1116 | - | - | - |
| $Cr_3C_2$ | 1160 | 1108 | - | - | - |

**Table 8b.** Same as Table 2b for WC phase in the non-rotating case at $10^{-6}$ bar pressure (WCNR_P6)

| Condensate | Static | Adiabatic with the non-equilibrium parameter, *f* | | | |
|---|---|---|---|---|---|
| **WCNR_P6** | ***f* = 1** | ***f* = 1** | **0.995** | **0.99** | **0.95** |
| C | $1.4\times10^{-1}$ | $1.4\times10^{-1}$ | $1.4\times10^{-1}$ | $1.4\times10^{-1}$ | $1.4\times10^{-1}$ |
| TiC | $2.8\times10^{-6}$ | $2.8\times10^{-6}$ | $2.8\times10^{-6}$ | $2.8\times10^{-6}$ | $2.8\times10^{-6}$ |
| SiC | $1.9\times10^{-4}$ | $2.5\times10^{-4}$ | $2.5\times10^{-4}$ | $1.1\times10^{-4}$ | $3.0\times10^{-5}$ |
| FeSi | $7.5\times10^{-4}$ | $7.5\times10^{-4}$ | $5.4\times10^{-4}$ | - | - |
| NiSi | $8.3\times10^{-5}$ | $8.2\times10^{-5}$ | $1.1\times10^{-4}$ | - | - |
| $Fe_3C$ | $5.0\times10^{-4}$ | $4.6\times10^{-4}$ | $5.6\times10^{-4}$ | - | - |
| CaS | $8.1\times10^{-5}$ | $7.8\times10^{-5}$ | $6.7\times10^{-5}$ | - | - |
| $CaAl_4O_7$ | $1.7\times10^{-5}$ | $1.1\times10^{-5}$ | - | - | - |
| $Cr_3C_2$ | $1.4\times10^{-5}$ | $5.8\times10^{-6}$ | - | - | - |

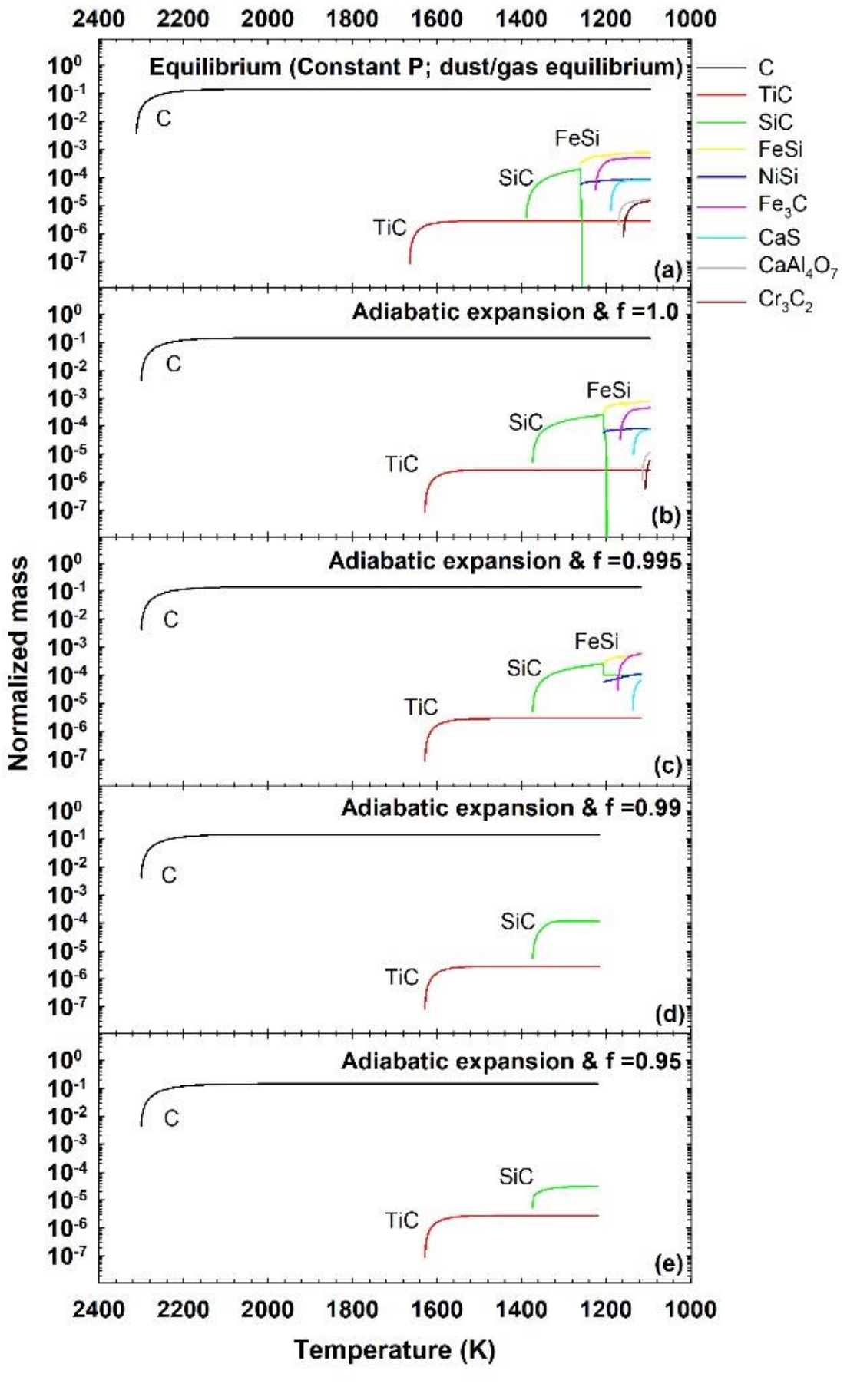

**Figure 9.** Same as Fig. 1 for WC composition in a non-rotating stellar model at a pressure of $10^{-6}$ bar. The major part of the dust mass is in the carbon (graphite) form.

**Table 9a.** Same as Table 2 for late-type WN phase at $10^{-6}$ bar pressure (WNL_P6)

| Condensate | Static | Adiabatic with the non-equilibrium parameter, *f* | | | |
|---|---|---|---|---|---|
| **WNL_P6** | ***f* = 1** | ***f* = 1** | **0.995** | **0.99** | **0.95** |
| TiC | 1583 | 1561 | 1561 | 1561 | 1561 |
| SiC | 1442 | 1412 | 1412 | 1412 | 1412 |
| (Fe, Ni) Si | 1228 | 1188 | 1188 | 1188 | 1188 |
| SiC (*) | - | - | - | - | 1163 |
| AlN | 1197 | 1154 | 1154 | 1154 | 1154 |
| $Cr_5Si_3$ | 1193 | 1150 | 1150 | 1150 | 1145 |
| Co | 1169 | 1126 | 1126 | 1126 | - |
| TiN | 1122 | 1075 | 1075 | 1075 | - |
| TiC (*) | 1121 | 1074 | 1074 | 1074 | - |
| CaS | 1113 | 1067 | 1067 | 1067 | - |
| $Si_2N_2O$ | 1099 | 1054 | 1054 | 1054 | - |

**Table 9b.** Same as Table 2b for late-type WN phase at $10^{-6}$ bar pressure (WNL_P6)

| Condensate | Static | Adiabatic with the non-equilibrium parameter, *f* | | | |
|---|---|---|---|---|---|
| **WNL_P6** | ***f* = 1** | ***f* = 1** | **0.995** | **0.99** | **0.95** |
| TiC | $1.1\times10^{-6}$ | $1.1\times10^{-6}$ | $1.1\times10^{-6}$ | $1.1\times10^{-6}$ | $1.1\times10^{-6}$ |
| SiC | $9.2\times10^{-5}$ | $9.1\times10^{-5}$ | $9.1\times10^{-5}$ | $9.1\times10^{-5}$ | $8.8\times10^{-5}$ |
| NiSi | $1.5\times10^{-6}$ | $1.5\times10^{-6}$ | $1.5\times10^{-6}$ | $1.5\times10^{-6}$ | $1.5\times10^{-6}$ |
| FeSi | $6.3\times10^{-6}$ | $6.3\times10^{-6}$ | $1.3\times10^{-5}$ | $1.3\times10^{-5}$ | $8.8\times10^{-6}$ |
| AlN | $1.4\times10^{-5}$ | $1.4\times10^{-5}$ | $1.4\times10^{-5}$ | $1.4\times10^{-5}$ | $1.0\times10^{-5}$ |
| $Cr_5Si_3$ | $7.3\times10^{-6}$ | $7.3\times10^{-6}$ | $7.3\times10^{-6}$ | $7.2\times10^{-6}$ | $1.2\times10^{-6}$ |
| TiN | $1.1\times10^{-6}$ | $1.1\times10^{-6}$ | $2.5\times10^{-8}$ | $8.3\times10^{-8}$ | - |
| CaS | $3.4\times10^{-5}$ | $3.4\times10^{-5}$ | $3.5\times10^{-5}$ | $2.9\times10^{-5}$ | - |
| $Si_2N_2O$ | $1.2\times10^{-4}$ | $1.2\times10^{-4}$ | $1.2\times10^{-4}$ | $4.7\times10^{-5}$ | - |

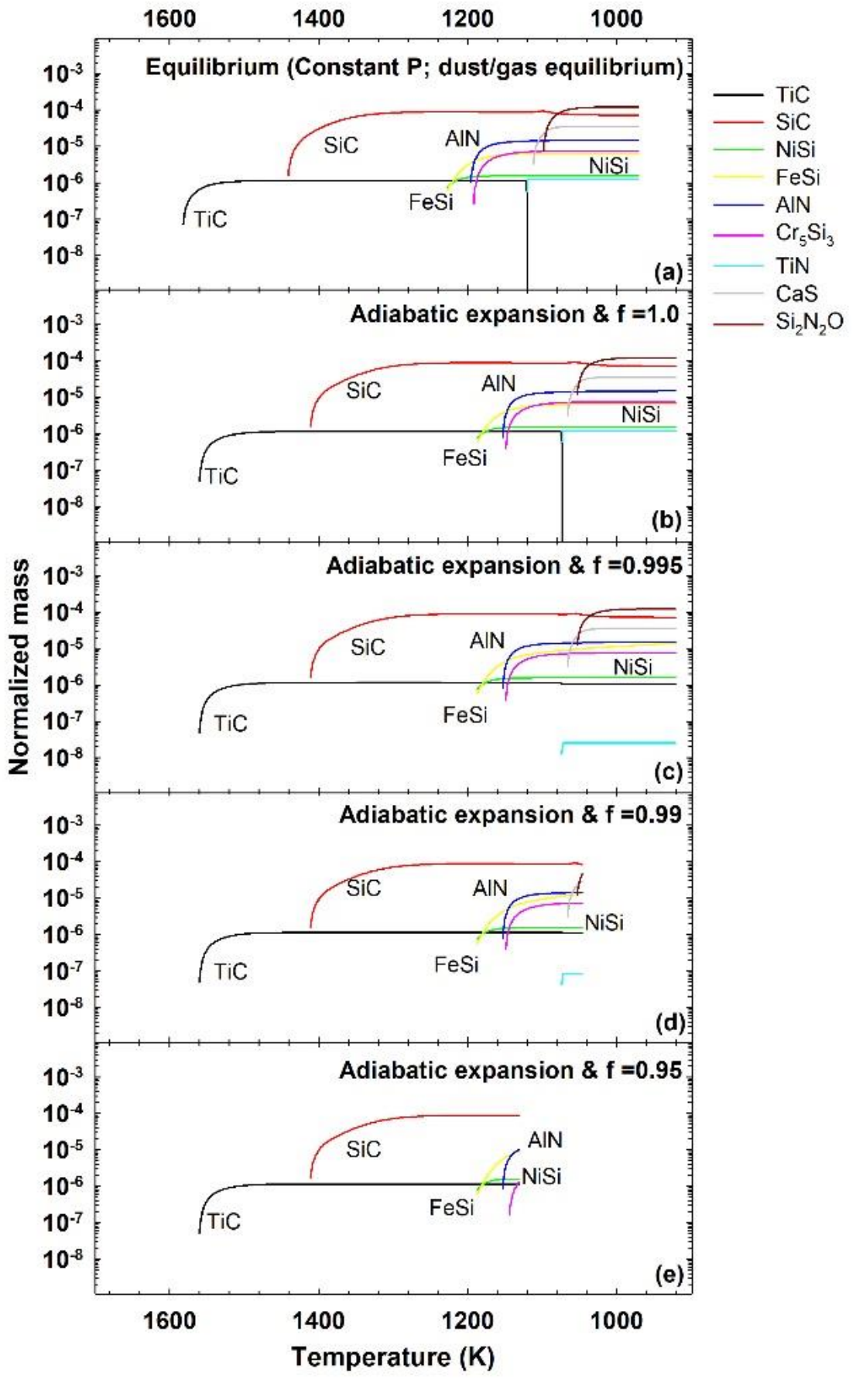

**Figure 10.** Same as Fig. 1 but for a late-type WN phase at a pressure of $10^{-6}$ bar. The major part of the dust mass is in carbide or nitride form.

The influence of pressure on the condensation of dust grains can be clearly inferred from the simulations performed in the present work. The panel (a) in these figures (Figs. 1, 2, 5-10) represent thermodynamical scenario at constant total pressure, whereas, the panels (b) represent the scenarios involving dynamic pressure in the cooling and expanding WR winds. The results in the scenario (b) are almost identical to the scenario (a) but the condensation temperatures of all the condensates were observed to reduce with the decrease in pressure. Similarly, for the second scenario, the lower value of system total pressure shifted the stability fields of the condensates downwards (Figs. 2a & 9a and 7a & 10a). The stability field gets affected by the system pressure but this window of the change in the stability field with the change in pressure is different for all species. The chemistry of all the species in the assemblage is coupled. In general, the appearance of the condensates in the system showed a similar trend, but in a few cases, the change in the stability field of condensates with the change in pressure may produce significant changes in the involved reactions, and hence, the condensation sequence. For instance, the stability field of SiC shifts upward in the case of the WC phase for a non-rotating stellar model at $10^{-6}$ bar total pressure (Fig. 9; Table 8). The numerical code does not infer any other major difference among the simulated models with distinct system pressure. The condensation sequence also remains almost the same for adiabatic variation in pressure.

## 5 CONCLUSION

In the present study, we have performed thermodynamic equilibrium as well as non-equilibrium calculations for dust condensation in various Wolf-Rayet phases of very massive stars. The mass distribution of the condensates, the condensation reactions and the condensation sequence have been computed for various WR stellar compositions. The significant differences in the relative elemental abundance affect the thermodynamics and the equilibria chemistry of the involved species. The initial chemical composition and the system pressure are the main parameters to study the condensation of dust grains in any cooling astrophysical environment. We have performed numerical simulations for a wide variety of chemical compositions of WR stars in WN and WC phases at various dust to gas equilibrium fractions with the system pressure static as well as dynamic in an adiabatic manner.

1. The condensation sequence for non-equilibrium scenario does show some distinctions from the equilibrium calculations. The major differences between the two scenarios can be seen in disappearance temperatures and the number of condensates stabilizing in the system which depend upon the non-equilibrium thermodynamical parameter.
2. From the thermodynamical equilibrium as well as non-equilibrium condensation calculations, we infer the condensation of carbides, nitrides, and sulphides in various N-rich and C-rich evolutionary phases of the massive stellar ejecta.
3. The grains condensed in the circumstellar envelope of WR stars can also be a source of isotopic anomalies in the early solar system.
4. C, TiC, SiC, AlN, CaS and Fe-metal are the new suite of refractory minerals which appear for several variations in this reducing environment. These are the most likely solids to have formed in WR stars.
5. The dust mass was found to be less than 1 per cent of the total gas mass for the WN phase, whereas, higher dust mass has been predicted for the WC phase.
6. The dust mass is observed to be less than 1 per cent on the basis of observation of WC star (Williams 1995). The dust mass agrees with this value for very initial stages of condensation in case of WC stars at $10^{-6}$ bar pressure.

## ACKNOWLEDGEMENTS

We are extremely grateful for the numerous comments and suggestions made by the reviewer that led to significant improvement of the article. AG acknowledges the Council of Scientific & Industrial Research (CSIR) for providing the doctoral financial assistance (# 09/135(0716)/2015-EMR-I) using which the present code has been written. AG and SS acknowledge the funding from the Indian Space Research Organisation (ISRO) through PLANEX grant for providing and maintaining the theoretical laboratory facilities.

# Supplementary Data

Table S1. Library of all the considered species and sources of their thermodynamic data.

**Gas Species**

Al, $Al_2$, $Al_2Cl_6$, $Al_2F_6$, $Al_2O$, $Al_2O_2$, AlC, AlCl, $AlCl_2$, $AlCl_2F$, $AlCl_3$, AlClF, $AlClF_2$, AlF, $AlF_2$, $AlF_3$, AlH, AlN, AlO, $AlO_2$, $AlO_2H$, AlOH, AlS, C, $C_2$, $C_2Cl_2$, $C_2Cl_4$, $C_2Cl_6$, $C_2F_2$, $C_2F_3N$, $C_2F_4$, $C_2F_6$, $C_2H$, $C_2H_2$, $C_2H_4$, $C_2H_4O$, $C_2HCl$, $C_2HF$, $C_2N$, $C_2N_2$, $C_2O$, $C_3$, $C_3O_2$, $C_4$, $C_4N_2$, $C_5$, Ca, $Ca_2$, CaCl, $CaCl_2$, CaF, $CaF_2$, CaO, $CaO_2H_2$, CaOH, CaS, CCl, $CCl_2$, $CCl_2F_2$, $CCl_3$, $CCl_3F$, $CCl_4$, $CClF_3$, CF, $CF_2$, $CF_3$, $CF_4$, $CF_4O$, $CF_8S$, CH, $CH_2$, $CH_2Cl_2$, $CH_2ClF$, $CH_2F_2$, $CH_3$, $CH_3Cl$, $CH_3F$, $CH_4$, CHCl, $CHCl_2F$, $CHCl_3$, $CHClF_2$, CHF, $CHF_3$, CHP, Cl, $Cl_2O$, $Cl_2S_2$, ClCN, ClF, $ClF_3$, $ClF_5$, ClO, $ClO_2$, $ClO_3F$, CN, $CN_2$, CO, Co, $CO_2$, $Co_2Cl_4$, COCl, CoCl, $COCl_2$, $CoCl_2$, $CoCl_3$, COClF, COF, $COF_2$, $CoF_2$, COS, CP, Cr, CrN, CrO, $CrO_2$, $CrO_3$, CS, $CS_2$, F, $F_2$, $F_2S_2$, FCN, Fe, $Fe_2Cl_4$, $Fe_2Cl_6$, $FeC_5O_5$, FeCl, $FeCl_2$, $FeCl_3$, FeF, $FeF_2$, $FeF_3$, FeO, $FeO_2H_2$, FeS, $FH_3Si$, FHO, $FHSO_3$, $F_2N_2$, FNO, $FNO_2$, $FNO_3$, FP, FPS, H, $H_2$, $H_2CO$, $H_2F_2$, $H_2O$, $H_2O_2$, $H_2S$, $H_2SO_4$, $H_3F_3$, $H_4F_4$, $H_5F_5$, $H_6F_6$, $H_7F_7$, HCl, HCN, HCO, HCOF, He, HF, HNCO, HNO, $HNO_3$, $HO_2$, HOCl, $HNO_2$, HS*, K, $K_2$, $K_2C_2N_2$, $K_2Cl_2$, $K_2F_2$, $K_2O_2H_2$, $K_2SO_4$, KCl, KCN, KF, KH, KO, KOH, Mg, $Mg_2$, $Mg_2Cl_4$, $Mg_2F_4$, MgCl, $MgCl_2$, MgClF, MgF, $MgF_2$, MgH, MgN, MgO, $Mg(OH)_2$, MgOH, MgS, N, $N_2$, $N_2F_4$, $N_2H_2$, $N_2H_4$, $N_2O$, $N_2O_3$, $N_2O_4$, $N_2O_5$, $N_3$, Na, $Na_2$, $Na_2(CN)_2$, $Na_2Cl_2$, $Na_2F_2$, $Na_2(OH)_2$, $Na_2SO_4$, $NaAlF_4$, NaCl, NaCN, NaF, NaH, NaO, NaOH, NCO, NF, $NF_2$, $NF_3$, $NF_3O$, NH, $NH_2$, $NH_3$, Ni, $NiC_4O_4$, NiCl, $NiCl_2$, NiS, NO, $NO_2$, $NO_2Cl$, $NO_3$, NS, O, $O_2$, $O_2F$, $O_3$, OAlCl, OAlF, $OAlF_2$, OF, $OF_2$, OH, ONCl, $OPCl_3$, $OSiF_2$, OTiCl, OTiF, $OTiF_2$, P, $P_2$, $P_4$, $P_4O_{10}$, $P_4O_6$, $P_4S_3$, PCl, $PCl_3$, $PCl_5$, $PF_2$, $PF_3$, $PF_5$, PH**, $PH_2$, $PH_3$**, PN**, PO, $PO_2$, $POCl_2F$, $POClF_2$, $POF_3$, PS, $PSF_3$, S, $S_2$, $S_2Cl$, $S_2F_{10}$, $S_2O$, $S_3$, $S_4$, $S_5$, $S_6$, $S_7$, $S_8$, SCl, $SCl_2$, $SClF_5$, SF, $SF_2$, $SF_3$, $SF_4$, $SF_5$, $SF_6$, Si, $Si_2$, $Si_2C$, $Si_2N$, $Si_3$, SiC, $SiC_2$, $SiC_4H_{12}$, $SiCH_3Cl_3$, $SiCH_3F_3$, SiCl, $SiCl_2$, $SiCl_3$, $SiCl_3F$, $SiCl_4$, $SiClF_3$, SiF, $SiF_2$, $SiF_3$, $SiF_4$, SiH, $SiH_2Cl_2$, $SiH_2F_2$, $SiH_3Cl$, $SiH_4$, $SiHCl_3$, $SiHF_3$, SiN, SiO, $SiO_2$, SiS, SO, $SO_2$, $SO_2Cl_2$, $SO_2ClF$, $SO_2F_2$, $SO_3$, $SPCl_3$, Ti, TiCl, $TiCl_2$, $TiCl_3$, $TiCl_4$, TiF, $TiF_2$, $TiF_3$, $TiF_4$, TiO, $TiO_2$, $TiOCl_2$

**Liquid Species**

Al, $Al_2O_3$, $AlCl_3$, $AlF_3$, Ca, $CaCl_2$, $CaF_2$, CaO, ClSSCl, Co, $CoCl_2$, $CoF_2$, Cr, $Cr_2O_3$, Fe, $Fe(CO)_5$, $FeCl_2$, $FeCl_3$, $FeF_2$, FeO, FeS, $H_2O$, $H_3PO_4$, K, $K_2CO_3$, $K_2S$, $K_2SiO_3$, $K_2SO_4$, KCl, KCN, KF, $K(HF_2)$, KOH, Mg, $Mg_2Si$, $Mg_2SiO_4$, $Mg_2TiO_4$, $Mg_3P_2O_8$, $MgAl_2O_4$, $MgCl_2$, $MgF_2$, MgO, $MgSiO_3$, $MgSO_4$, $MgTi_2O_5$, $MgTiO_3$, $N_2H_4$, $N_2O_4$, Na, $Na_2CO_3$, $Na_2O$, $Na_2S$, $Na_2S_2$, $Na_2Si_2O_5$, $Na_2SiO_3$, $Na_2SO_4$, $Na_3AlF_6$, $Na_5Al_3F_{14}$, NaCl, NaCN, NaF, NaOH, Ni, $Ni_3S_2$, $Ni(CO)_4$, $NiCl_2$, NiS, $NiS_2$, P, $P_4S_3$, S, $SCl_2$, Si, $SiO_2$, $SiS_2$, Ti, $Ti_2O_3$, $Ti_3O_5$, $Ti_4O_7$, TiC, $TiCl_4$, TiN, TiO, $TiO_2$

**Solid Species*****

| | | |
|---|---|---|
| Al [1] | $Na_2O$ [1] | $CaAl_2SiO_6$ (Pyroxene) [2] |
| $Al_2O_3$ (alpha) [1] | $Na_2O_2$ [1] | $CaAl_4O_7$ (Grossite) [2] |
| $Al_2O_3$ (gamma) [1] | $Na_2S$ [1] | $CaC_2$ [2] |
| $Al_2O_3$ (kappa) [1] | $Na_2S_2$ (beta) [1] | $CaCO_3$ (Aragonite) [2] |
| $Al_2O_3$ (delta) [1] | $Na_2Si_2O_5$ [1] | $CaCO_3$ (Calcite) [2] |
| $Al_2S_3$ [1] | $Na_2SiO_3$ [1] | $CaFe_2O_4$ [2] |
| $Al_2SiO_5$ (andalusite) [1] | $Na_2SO_4$ (I) [1] | $CaH_2$ [2] |
| $Al_2SiO_5$ (kyanite) [1] | $Na_2SO_4$ (III) [1] | $CaHPO_4$ [2] |
| $Al_2SiO_5$ (sillimanite) [1] | $Na_2SO_4$ (IV) [1] | $CaMg_2$ [2] |
| $Al_4C_3$ [1] | $Na_2SO_4$ (delta) [1] | $CaMgO_2$ [2] |
| $Al_6Si_2O_{13}$ (mullite) [1] | $Na_2SO_4$ (V) [1] | $CaMgSi_2O_6$ (Diopside) [2] |
| $AlCl_3$ [1] | $Na_3AlCl_6$ [1] | $CaMgSiO_4$ (Monticellite) [2] |
| $AlF_3$ [1] | $Na_3AlF_6$ (alpha) [1] | CaSi [2] |
| AlN [1] | $Na_3AlF_6$ (beta) [1] | $CaSi_2$ [2] |
| Ca (alpha) [1] | $Na_5Al_3F_{14}$ [1] | $CaSiO_3$ (Wollastonite) [2] |
| Ca (beta) [1] | $NaAlCl_4$ [1] | $CaSiO_3$ (Pseudowollastonite) [2] |
| $CaCl_2$ [1] | $NaAlO_2$ [1] | $CaSO_4$ (Anhydrite) [2] |
| $CaF_2$ [1] | NaCl [1] | $CaTiO_3$ (Perovskite) [2] |
| CaO [1] | $NaClO_4$ [1] | $CaTiSiO_5$ (Sphene) [2] |
| $Ca(OH)_2$ [1] | NaCN [1] | $Co_2SiO_4$ [2] |
| CaS [1] | NaF [1] | $CoS_2$ [2] |

$Co$ [1]
$Co_3O_4$ [1]
$CoCl_2$ [1]
$CoF_2$ [1]
$CoF_3$ [1]
$CoO$ [1]
$CoSO_4$ [1]
$Cr$ [1]
$Cr_{23}C_6$ [1]
$Cr_2N$ [1]
$Cr_2O_3$ [1]
$Cr_3C_2$ [1]
$Cr_7C_3$ [1]
$CrN$ [1]
Fe (gamma) [1]
$Fe_2O_3$ (hematite) [1]
$Fe_2(SO_4)_3$ [1]
$Fe_3O_4$ (magnetite) [1]
$FeCl_2$ [1]
$FeCl_3$ [1]
$FeF_2$ [1]
$FeF_3$ [1]
$FeO$ [1]
$Fe(OH)_2$ [1]
$Fe(OH)_3$ [1]
FeS (troilite) [1]
$FeS_2$ (marcasite) [1]
$FeS_2$ (pyrite) [1]
$FeSO_4$ [1]
$H_3PO_4$ [1]
K [1]
$K_2CO_3$ [1]
$K_2O$ [1]
$K_2O_2$ [1]
$K_2S$ [1]
$K_2SiO_3$ [1]
$K_2SO_4$ (alpha) [1]
$K_2SO_4$ (beta) [1]
$K_3Al_2Cl_9$ [1]
$K_3AlCl_6$ [1]
$K_3AlF_6$ [1]
$KAlCl_4$ [1]
KCl [1]
$KClO_4$ [1]
KCN [1]
KF [1]
KH [1]
$K(HF_2)$ [1]
NaH [1]
$NaO_2$ [1]
NaOH [1]
$NH_4Cl$ [1]
$NH_4ClO_4$ [1]
Ni [1]
$Ni_3S_2$ [1]
$Ni_3S_4$ [1]
$NiCl_2$ [1]
NiS [1]
$NiS_2$ [1]
OAlCl [1]
P (white) [1]
P (black) [1]
P (red_IV) [1]
P (red_V) [1]
$P_3N_5$ [1]
$(P_2O_5)_2$ [1]
$P_4S_3$ [1]
S (orthorhombic) [1]
S (monoclinic) [1]
Si [1]
$Si_3N_4$ (alpha) [1]
SiC (alpha) [1]
SiC (beta) [1]
$SiO_2$ (cristobalite Low) [1]
$SiO_2$ (cristobalite High) [1]
$SiO_2$ (quartz) [1]
$SiS_2$ [1]
Ti (alpha) [1]
Ti (beta) [1]
$Ti_2O_3$ [1]
$Ti_3O_5$ (alpha) [1]
$Ti_3O_5$ (beta) [1]
$Ti_4O_7$ [1]
TiC [1]
$TiCl_2$ [1]
$TiCl_3$ [1]
$TiCl_4$ [1]
$TiF_3$ [1]
$TiF_4$ [1]
$TiH_2$ [1]
TiN [1]
TiO (alpha) [1]
TiO (beta) [1]
$TiO_2$ (anatase) [1]
$TiO_2$ (rutile) [1]
$Al_2TiO_5$ [2]
$Cr_3Si$ [2]
$Cr_5Si_3$ [2]
$CrCl_2$ [2]
$CrCl_3$ [2]
$CrO_2$ [2]
$CrO_3$ [2]
CrS [2]
CrSi [2]
$CrSi_2$ [2]
$Fe_2SiO_4$ (Fayalite) [2]
$Fe_2TiO_4$ [2]
$Fe_3C$ (Cohenite) [2]
$Fe_4N$ [2]
$FeAl_2O_4$ (Hercynite) [2]
$FeCO_3$ [2]
FeOCl [2]
FeSi [2]
$FeSi_2$ [2]
$FeSiO_3$ (Ferrosilite) [2]
$FeTiO_3$ (Ilmenite) [2]
$K_2CrO_4$ [2]
$K_2HPO_4$ [2]
$K_2Si_2O_5$ [2]
$K_2Si_4O_9$ [2]
$K_3PO_4$ [2]
$KAlSi_2O_6$ [2]
$KAlSi_3O_8$ (Adularia) [2]
$KAlSi_3O_8$ (Microcline) [2]
$KAlSi_3O_8$ (Sanidine) [2]
$KAlSiO_4$ [2]
$KH_2PO_4$ [2]
$KNO_3$ [2]
$Mg_2Al_4Si_5O_{18}$ (Cordierite) [2]
$Mg_3Al_2Si_3O_{12}$ (Pyrope) [2]
$Mg_3Si_4O_{12}H_2$ (Talc) [2]
$Mg_7Si_8O_{24}H_2$ (Anthophyllite) [2]
$MgFe_2O_4$ [2]
$MgNi_2$ [2]
Mg(OH)Cl [2]
$Na_2CrO_4$ [2]
$Na_2S_3$ [2]
$Na_2S_4$ [2]
$Na_2Ti_2O_5$ [2]
$Na_2Ti_3O_7$ [2]
$Na_2TiO_3$ [2]
$Na_3PO_4$ [2]
$Na_4SiO_4$ [2]
$Na_6Si_2O_7$ [2]

| | | |
|---|---|---|
| $KO_2$ [1] | AlOCl [2] | $NaAlSi_2O_6$ (Jadeite) [2] |
| KOH [1] | AlP [2] | $NaAlSi_2O_6$ (Dehydrated Analcime) [2] |
| Mg [1] | $AlPO_4$ [2] | $NaAlSi3O_8$ (Albite) [2] |
| $Mg_2C_3$ [1] | C (Graphite) [2] | $NaAlSiO_4$ (Nepheline) [2] |
| $Mg_2Si$ [1] | C (Diamond) [2] | $NaNO_3$ [2] |
| $Mg_2SiO_4$ [1] | $Ca_{12}Al_{14}O_{33}$ [2] | $Ni_2P$ [2] |
| $Mg_2TiO_4$ [1] | $Ca_2Al_2O_5$ [2] | $Ni_2SiO_4$ [2] |
| $Mg_3N_2$ [1] | $Ca_2Al_2SiO_7$ (Gehlenite) [2] | $Ni_3C$ [2] |
| $Mg_3P_2O_8$ [1] | $Ca_2Fe_2O_5$ [2] | $Ni_3P$ [2] |
| $MgAl_2O_4$ [1] | $Ca_2MgSi_2O_7$ (Akermanite) [2] | $Ni_3Ti$ [2] |
| $MgC_2$ [1] | $Ca_2P_2O_7$ [2] | $Ni_5P_2$ [2] |
| $MgCl_2$ [1] | $Ca_2Si$ [2] | $Ni_7Si_{13}$ [2] |
| $MgCO_3$ [1] | $Ca_2SiO_4$ (Ca-Olivine) [2] | $NiAl_2O_4$ [2] |
| $MgF_2$ [1] | $Ca_2SiO_4$ (Larnite) [2] | $NiCO_3$ [2] |
| $MgH_2$ [1] | $Ca_3Al_2O_6$ [2] | NiO [2] |
| MgO [1] | $Ca_3Al_2Si_3O_{12}$ (Grossular) [2] | NiSi [2] |
| $Mg(OH)_2$ [1] | $Ca_3MgSi_2O_8$ (Merwinite) [2] | NiTi [2] |
| MgS [1] | $Ca_3N_2$ [2] | $NiTi_2$ [2] |
| $MgSiO_3$ [1] | $Ca_3P_2$ [2] | $NiTiO_3$ [2] |
| $MgSO_4$ [1] | $Ca_3Si_2O_7$ (Rankinite) [2] | SiP [2] |
| $MgTi_2O_5$ [1] | $Ca_3SiO_5$ [2] | TiS [2] |
| $MgTiO_3$ [1] | $Ca_3Ti_2O_7$ [2] | $TiS_2$ [2] |
| $N_2O_4$ [1] | $Ca_4Ti_3O_{10}$ [2] | $Si_2N_2O$ (Sinoite) [4] |
| Na [1] | $CaAl_2O_4$ (Dmitriyivanovite) [2] | $CaAl_{12}O_{19}$ (Hibonite) [2,5] |
| $Na_2CO_3$ [1] | $CaAl_2Si_2O_8$ (Anorthite) [2] | |

**Solid-Solutions**

| | | | |
|---|---|---|---|
| Melilite: | $Ca_2Al_2SiO_7$-$Ca_2MgSi_2O_7$ | Fassaite: | $CaMgSi_2O_6$-$CaAl_2SiO_6$ |
| Al - Spinel: | $MgAl_2O_4$-$FeAl_2O_4$ | Olivine: | $Mg_2SiO_4$-$Fe_2SiO_4$ |
| Metal-alloy: | Fe-Ni | Clinopyroxene: | $MgSiO_3$-$FeSiO_3$ |
| Plagioclase: | $CaAl_2Si_2O_8$-$NaAlSi_3O_8$ | Ferro-Nickel Silicon | (Fe, Ni) Si |

---

**Sources:**

* HS gas is taken from source [2].
** PH, $PH_3$, PN have been taken from source [3]. All other gas species and liquids have been adopted from source [1].
*** The format followed in the table for the solid species is: Species Formula (Species Name) [Species Source of Thermodynamic Data].

**Semi-Analytical Approach:**
**We have given semi-analytical approach in the manuscript. The assumptions and the deduced equations have already been presented in that. Here, we present the step by step derivation of the final equation. The steps which have only been written here, are colored in red. The text, colored in green, has been copied from the manuscript and provided here just for the sake of completeness.**

In the semi-analytical approach, we assume that $N_e$ is the amount of an element, say X, present in the gas phase at any temperature T. With the decrease in temperature, the element condenses into a solid species. The change in the abundance in the gas phase with the change in temperature depends upon the abundance of an element at that temperature. Mathematically,

$$\frac{dN_e}{dT} = K_e N_e \qquad (6)$$

$K_e$ is a proportionality factor, and it determines the rate of change of $N_e$. It is assumed to be a constant, even though, it could be a function of temperature. This assumption will not influence our adopted numerical approach, whereby, we piecewise handle the non-equilibrium thermodynamics at every temperature drop in a systematic manner. A positive sign is taken because $N_e$ is decreasing with a decrease in temperature. Integrating, the equation (6) within the limits, we get an approximate solution.

$$\int_{N_{e0}}^{N_e} \frac{dN_e}{N_e} = \int_{T_c}^{T} K_e dT \qquad (7)$$

Here, $N_{e0}$, is the initial abundance of the element in the gaseous state at the earliest condensation temperature $T_C$ where the element, X, begins to condense.

$$\ln[N_e]_{N_{e0}}^{N_e} = K_e[T]_{T_c}^{T} \qquad (7a)$$

$$\ln\left(\frac{N_e}{N_{e0}}\right) = K_e(T - T_c) \qquad (7b)$$

The equation (7) yields an analytical solution, presented in equation (8), for the abundance of the element, X, in the gas phase at any falling temperature step.

$$N_e = N_{e0}\, e^{K_e(T-T_c)} \qquad (8)$$

The element, X, can be a constituent of several condensates. The mass of condensate increases with the decrease in temperature as the condensation process proceeds. A condensate can further react with the gas, and thus, its abundance can decrease with the formation of new condensates. The change in the amount ($N_c$) of the element, X, in condensate on account of condensation as well as conversion to another distinct dust species can be written as,

$$\frac{dN_c}{dT} = -f_c K_e N_e + K_c N_c \qquad (9)$$

The first term here represents the condensation of an element in one specific dust species, whereas, the second term deals with the conversion of the dust species into another form on account of the gas-dust chemical reaction. $K_c$ is a constant proportionality factor. It determines the rate of change of the first condensed species on account of conversion to another dust species. $f_c$ is the fraction of the element, X, that condenses into the first condensate, and its value is $\leq 1$ as one element can condense in more than one condensate. However, if we assume the conversion into only one another condensate, we can obtain a simple analytical solution for the equation (9). Multiplying both sides by $e^{-K_c T}$ and rearranging the terms, we get

$$dN_c - K_c N_c dT = -f_c K_e N_e dT \qquad (9a)$$

$$(dN_c)e^{-K_c T} - K_c N_c e^{-K_c T} dT = -f_c K_e N_e e^{-K_c T} dT \qquad (9b)$$

Substituting the value of $N_e$ from equation (3), equation becomes

$$d(N_c e^{-K_c T}) = -f_c K_e N_{e0}\, e^{K_e(T-T_c)} e^{-K_c T} dT \qquad (9c)$$

Integrating 9c equation within the limits, we get

$$\int_{T_c}^{T} d(N_c e^{-K_c T}) = -f_c K_e N_{e0}\, e^{-K_e T_C} \int_{T_c}^{T} e^{(K_e - K_c)T}\, dT \tag{9d}$$

Assuming $N_c = 0$ at $T = T_c$,

$$N_c e^{-K_c T} = -f_c K_e N_{e0}\, e^{-K_e T_c} \left[\frac{e^{(K_e - K_c)T}}{(K_e - K_c)}\right]_{T_c}^{T} \tag{9e}$$

$$N_c e^{-K_c T} = -\frac{f_c K_e N_{e0}\, e^{-K_e T_c}}{(K_e - K_c)} \left[e^{(K_e - K_c)T} - e^{(K_e - K_c)T_c}\right] \tag{9f}$$

$$N_c = \frac{-f_c K_e N_{e0}}{(K_e - K_c)} \left[e^{K_e(T - T_C)} - e^{K_c(T - T_C)}\right] \tag{10}$$

The condensate abundance decreases with the formation of other dust species. Further, we assume that some fraction of the condensate always gets isolated from the gaseous system in terms of gas-dust non-equilibrium, as mentioned earlier. This non-equilibrium content of the condensate gradually increases according to the abundance of the condensate. The gradual increase in the non-equilibrium fraction of the element, X, due to the segregation of the dust from the gas component can be described in terms of $N_{ic}$, according to equation (11).

$$\frac{dN_{ic}}{dT} = -f_{ic} K_c N_c \tag{11}$$

Here, $f_{ic}$ is the part of the condensate that got isolated from the gaseous system due to the non-equilibrium dust-gas system. The equation (11) can be analytically solved using the initial abundance, $N_{ic}$, of the condensate for the element, X, obtained from the equation (10).